\documentclass[journal]{IEEEtran}
\usepackage[T1]{fontenc}

\usepackage{cite}
\usepackage{amsmath,graphicx}
\usepackage{amssymb}
\usepackage{amsfonts}
\usepackage{graphicx,subfigure}
\usepackage{fancyhdr}  %
\usepackage{cases}
\usepackage{extarrows}
\usepackage{algorithm}
\usepackage{algorithmic}
\usepackage{multirow,tabularx}
\usepackage{mathtools}
\usepackage{xcolor}
\usepackage[english]{babel}
\usepackage{caption}
\usepackage{bm}
\usepackage{graphicx}
\usepackage{epstopdf}

\newtheorem{lemma}{Lemma}
\newtheorem{remark}{Remark}

\def\proof{\noindent\hspace{2em}{\itshape Proof: }}
\def\endproof{\hspace*{\fill}~$\square$\par\endtrivlist\unskip}

\begin{document}
	\title{Bayesian Learning for Double-RIS Aided ISAC Systems with Superimposed Pilots and Data}
	\author{Xu Gan, Chongwen Huang, Zhaohui Yang, Caijun Zhong, Xiaoming Chen, Zhaoyang Zhang, Qinghua Guo, \\ Chau Yuen,~\IEEEmembership{Fellow,~IEEE}, M\'{e}rouane Debbah,~\IEEEmembership{Fellow,~IEEE}
		\thanks{X. Gan, C. Huang, Z. Yang, C. Zhong, X. Chen and Z. Zhang are with the Institute of Information and Communication Engineering, Zhejiang University, Hangzhou 310027,  (e-mails: \{gan\_xu, chongwenhuang, yang\_zhaohui, caijunzhong, chen\_xiaoming, zhzy\}@zju.edu.cn).
			
			Qinghua Guo is with the School of Electrical, Computer and Telecommunications Engineering, University of Wollongong, (e-mail: qguo@uow.edu.au).
			
			C. Yuen is with School of Electrical and Electronics Engineering, Nanyang Technological University, Singapore 639798 (e-mail: chau.yuen@ntu.edu.sg).
			
			M. Debbah is with Khalifa University of Science and Technology, P O Box 127788, Abu Dhabi, UAE (email: merouane.debbah@ku.ac.ae).}
		
	}

	\maketitle
	
	\begin{abstract}
		Reconfigurable intelligent surface (RIS) has great potential to improve the performance of integrated sensing and communication (ISAC) systems, especially in scenarios where line-of-sight paths between the base station and users are blocked. However, the spectral efficiency (SE) of RIS-aided ISAC uplink transmissions may be drastically reduced by the heavy burden of pilot overhead for realizing sensing capabilities. In this paper, we tackle this bottleneck by proposing a superimposed symbol scheme, which superimposes sensing pilots onto data symbols over the same time-frequency resources. Specifically, we develop a structure-aware sparse Bayesian learning framework, where decoded data symbols serve as side information to enhance sensing performance and increase SE. To meet the low-latency requirements of emerging ISAC applications, we further propose a low-complexity simultaneous communication and localization algorithm for multiple users. This algorithm employs the unitary approximate message passing in the Bayesian learning framework for initial angle estimate, followed by iterative refinements through reduced-dimension matrix calculations. Moreover, the sparse code multiple access technology is incorporated into this iterative framework for accurate data detection which also facilitates localization. Numerical results show that the proposed superimposed symbol-based scheme empowered by the developed algorithm can achieve centimeter-level localization while attaining up to $96\%$ of the SE of conventional communications without sensing capabilities. Moreover, compared to other typical ISAC schemes, the proposed superimposed symbol scheme can provide an effective throughput improvement over $133\%$.
		
		\begin{center}
			{\bf Index Terms}
		\end{center}
		Integrated sensing and communication (ISAC), reconfigurable intelligent surface (RIS), superimposed coding, Bayesian learning, unitary approximate message passing (UAMP).
	\end{abstract}

	\section{Introduction}
	Next-generation wireless networks have been envisioned as key enablers to provide integrated sensing and communication (ISAC) capabilities \cite{ISAC1,ISAC2} to support several environment- and location-aware intelligent applications that strongly rely on ultra-reliable sensing techniques. With this vision in mind, ISAC is designed to exploit the dual use of radio signals and hardware infrastructures for both wireless sensing and communications, and has attracted growing research interests in academia and industry \cite{ISAC4}. The large bandwidth at millimeter wave (mmWave) frequencies enables higher data rates and sensing resolution, thereby making mmWave signals attractive to be incorporated in ISAC systems \cite{mm_ISAC}. However, operating at mmWave frequencies is quite challenging, mainly due to the severe propagation loss that often hinders line-of-sight (LoS) paths and prevents reliable ISAC connections.
	
	Fortunately, an emerging technology named reconfigurable intelligent surface (RIS) has been viewed as a promising solution to resolve this issue \cite{RIS1}. By properly adjusting phase shifts of digitally-controlled reflecting elements, the RIS can favorably reconfigure the wireless propagation environment to create virtual LoS (VLoS) paths \cite{RIS2}. Specifically, RIS is a fully passive two-dimensional array with many sub-wavelength controllable elements that can be tuned to change the physical properties of the impinging electromagnetic wave flexibly. Moreover, the RIS is generally fabricated with inexpensive hardware components without radio-frequency (RF) chains \cite{RIS3}, thereby significantly reducing hardware cost and energy consumption. Up to now, plenty of research has been devoted to the potential applications of RIS-aided ISAC systems \cite{RIS_ISAC1,RIS_ISAC2,RIS_ISAC3,RIS_ISAC4,RIS_ISAC5}. In particular, the theoretical ISAC performance gain offered by RISs was quantitatively given in \cite{RIS_ISAC3}, which also derived optimal RIS designs in different scenarios. Besides, authors in \cite{RIS_ISAC4} and \cite{RIS_ISAC5} tuned the RIS to enhance the coupling degree of ISAC channels by modifying the underlying subspace, which greatly enhanced ISAC performance. 
	
	Among all ISAC proposals for 6G networks, one practical and promising approach is to leverage the existing communication hardware infrastructures and explore advanced sensing signal processing technologies \cite{perceptive_mobile}, which averts the extra deployment of sensory networks. This new framework brings a research upsurge, but also introduces great challenges, especially for uplink ISAC scenarios. This is because sensing capabilities cannot be fulfilled by uplink communication waveforms directly. Hence, dedicated sensing pilots are needed, which will consume additional time-frequency resources \cite{RIS_ISAC_time} and impair spectral efficiency (SE). Furthermore, accurately locating multiple user equipments (UEs) remains a challenging task, even based on pure sensing pilots in double-RIS aided systems. This is because the implementation of multi-UE localization is achieved by extracting channel angles of each UE, while they are intricately coupled in double-RIS cascaded channels. The same challenge exists in channel estimation of double-RIS aided systems, and has been studied in \cite{twoRIS2,twoRIS3,twoRIS4,hu_tsp}. In particular, authors in \cite{twoRIS2} proposed cascaded channel estimation methods for multi-user systems. Nevertheless, these methods result in a high amount of training complexities, and obtaining channel angles of UEs remains unresolved. These works \cite{twoRIS3,hu_tsp,twoRIS4} investigated this problem by using active RIS architectures with RF chains that have the ability to receive signals at the RIS. However, the deployment of active RISs increases hardware costs and energy consumption.
	
	In this paper, instead of allocating dedicated time-frequency resources for sensing functions or with active energy-consuming RISs, we propose a superimposed symbol scheme in double-RIS aided uplink ISAC scenarios, where sensing pilots are superimposed with data symbols\cite{superimposed1,superimposed2,superimposed3} to greatly reduce training/localization costs, thereby improving SE. Specifically, the channel angles used for localization can help to recover the communication data, and in turn, the demodulated data can be re-used as sensing pilots to enhance the localization accuracy. This scheme enables both high SE communications and accurate localization. The main contributions of this paper are summarized as follows.
	\begin{itemize}
		\item We develop a superimposed symbol scheme in double-RIS aided uplink ISAC mmWave scenarios. Compared to existing sensing pilot-based schemes, ours can yield better SE performance since there are little dedicated time-frequency resources reserved for sensing pilots.
		
		\item We propose a structure-aware sparse Bayesian learning (SBL) framework to extract channel angles from intricately coupled channels, where decoded data can be used as side information to improve sensing performance and increase SE. The proposed algorithm exploits two-timescale channel properties, and enables concurrent communication and channel angle sensing. 
		
		\item We further propose a low-complexity algorithm to realize multi-UE simultaneous communication and localization. Specifically, the unitary approximate message passing (UAMP) is employed in this Bayesian learning framework to avoid high-complexity matrix inversion operations for initial estimates. Then, the estimates are refined iteratively through reduced-dimension matrix calculations. Moreover, the sparse code multiple access (SCMA) technology is incorporated into the iterative framework to suppress multi-UE interference for accurate data detection that also facilitates localization. 
		
		\item We numerically evaluate the performance of proposed algorithms in a 3D ISAC scenario. The results indicate that our proposed superimposed symbol-based scheme empowered by the developed algorithm can achieve centimeter-level localization while reaching up to $96\%$ of the SE of conventional communications without sensing capabilities. Besides, compared to other typical schemes with sensing capabilities, ours can also yield larger effective throughput. 
		
	\end{itemize}
	
	\section{System Model}\label{model}
	In this section, the system scenario, the transmission protocol, the channel model, and the signal model are presented, all of which are tailored to the double-RIS model to support the multi-UE simultaneous localization and communication.
	
\begin{figure}[ht]
		\centering{\includegraphics[scale=0.2]{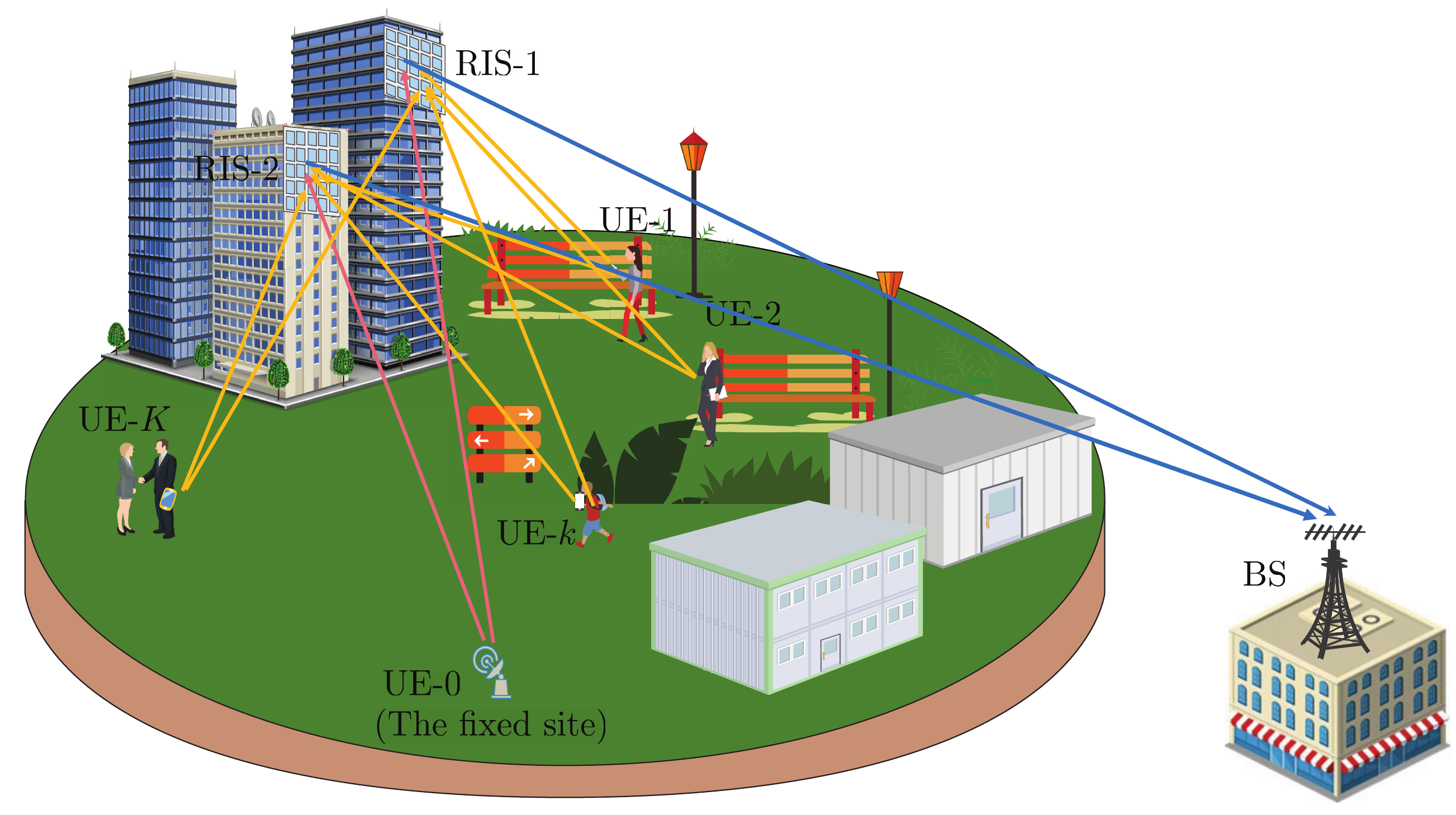}}\vspace{-2pt}
		\caption{Illustration of the double-RIS aided multi-UE uplink ISAC scenario.}\label{system}
	\end{figure}

	We consider a multi-UE uplink ISAC scenario in double-RIS aided mmWave systems as illustrated in Fig.~\ref{system}, where $K$ single-antenna UEs are served by a base station (BS) with the aid of RIS-$1$ and RIS-$2$ to enable simultaneous communication and localization. The BS has an $M$-antenna uniform linear array (ULA)\footnote{The proposed algorithm for ULA can be readily extended to the URA structure by introducing the azimuth angles of BS for estimation in a similar way.} along the $Y$-axis, while the $i$-th ($i\in\{1,2\}$) RIS has an $N_i = N_{i,\text{y}}\times N_{i,\text{z}}$ uniform rectangular array (URA) lying on the $Y$-o-$Z$ plane\footnote{Since the interference caused by double cascaded links in this setting can seriously impair the ISAC performance, we are opting for designing placement of these two RISs, i.e., both to be placed near UEs and have the same orientation, to eliminate this interference. }. Besides, it is assumed that the obstructions block LoS paths between the BS and UEs, which is common in mmWave scenarios. In light of this, RISs can be employed to provide favorable VLoS paths to attain satisfactory communication and localization performance. 
	
	\vspace{-10pt}
	\subsection{Transmission Protocol}
	We first design the transmission protocol to realize simultaneous communication and localization of multiple UEs in the considered system. The main challenge in implementing the localization functionality stems from extracting channel angles of each UE, which are intricately coupled with BS-RIS channels. Hence, prior knowledge of channel angle information for BS-RIS links is required. An efficient solution to sense BS-RIS channel angle information is to superimpose sensing pilots onto data symbols of a fixed site, such as a smart meter or camera, whose position is known prior. Then, based on the estimated BS-RIS channel angle information, the channel angles of each UE can be extracted for localization.
	
\begin{figure}[ht]
		\centering
		\includegraphics[scale=0.29]{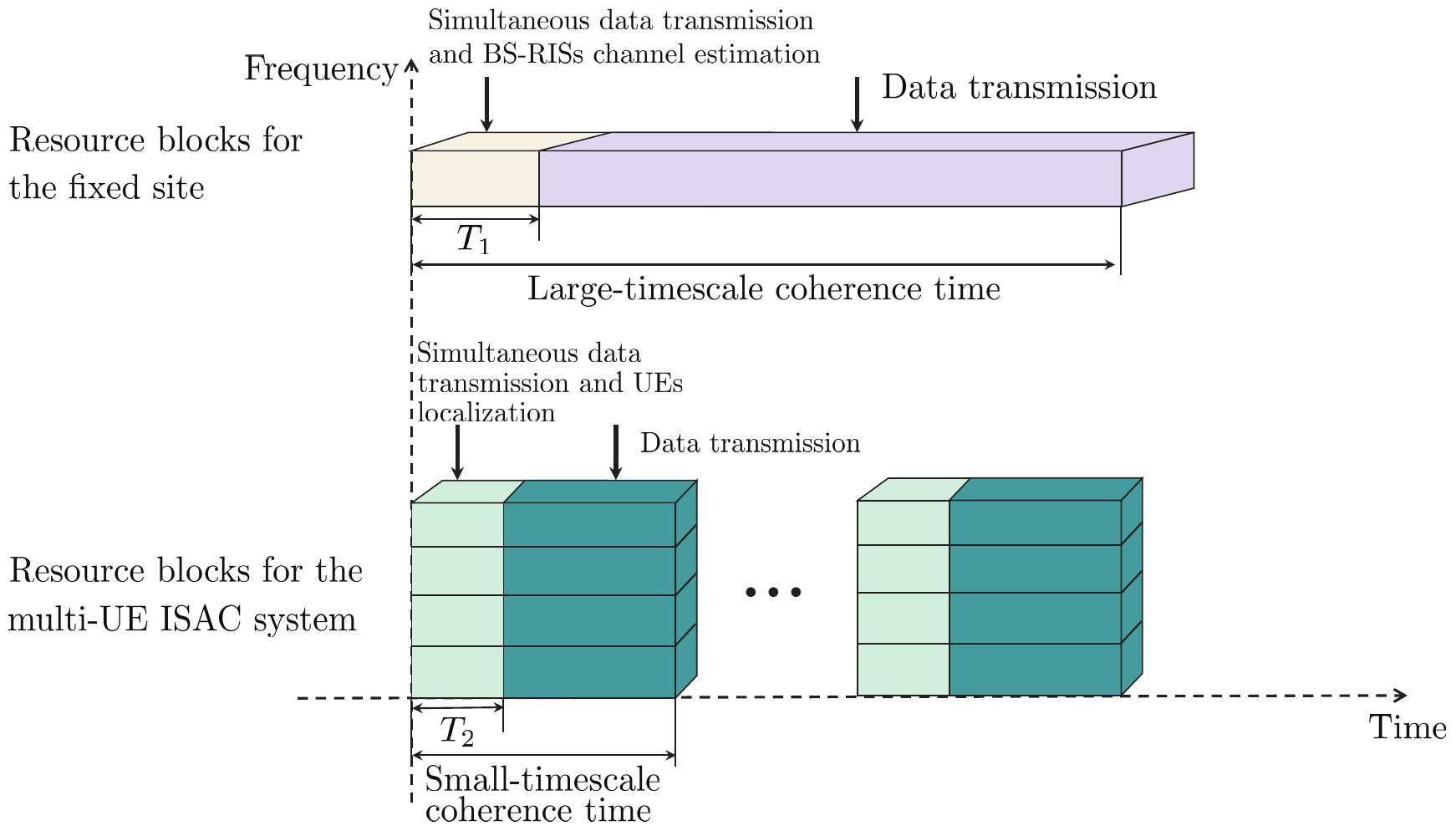}\vspace{-2pt}
		\caption{Illustration of two-timescale transmission protocol.}
		\vspace{-5pt}
		\label{protocol}
	\end{figure}

	Owing to the elevated fixed positions of the BS and RISs, channels between them are modeled as quasi-static, while channels between RISs and mobile UEs are fast time-varying. We can exploit these two-timescale channel properties and design the transmission protocol as shown in Fig.~\ref{protocol} to achieve multi-UE simultaneous communication and localization. Specifically, the fixed site transmits $T_1$ superimposed sensing pilots and data at the beginning of the large-timescale coherence time enabling BS-RIS channel sensing without interrupting the communication procedure. For frequent localization of multiple UEs, they need to transmit superimposed symbols at the beginning of every small-timescale coherent time within the $T_2$ period.

	\subsection{Channel Model}
	The channel between the BS and the $i$-th RIS can be characterized by the widely used Saleh-Valenzuela model \cite{mm_ch} due to the limited scattering characteristics, expressed as
	\begin{equation}\label{Hr}
		\mathbf{H}_{\text{r},i} = \sqrt{\frac{M N_i}{L_i}}\sum\limits_{l=1}^{L_i} \alpha_{l,i} \mathbf{a}_{M}(u_{\text{\text{R2B}},il}^{\text{A}}) \mathbf{b}_{N_i}^H(u_{\text{\text{R2B}},il}^{\text{D}}, v_{\text{\text{R2B}},il}^{\text{D}}),
	\end{equation}
	where $L_i$ represents the number of paths between the BS and the $i$-th RIS, and $\alpha_{l,i}$ represents the complex gain consisting of path loss for the $l$-th path. The array responses can be expressed as $\mathbf{a}_{M}(u) = \frac{1}{\sqrt{M}}[1,...,e^{j\pi(M-1)u}]^T$ and $\mathbf{b}_{N_i}(u,v) = \frac{1}{\sqrt{N_i}}[1,...,e^{j\pi(N_{i,\text{y}}-1)u}]^T \otimes [1,...,e^{j\pi(N_{i,\text{z}}-1)v}]^T$ with the effective angles of arrival (AoAs) and azimuth angles of departure (AoDs) as $u_{\text{\text{R2B}},il}^{\text{A}} = \frac{2d_{\text{BS}}}{\lambda} \sin(\vartheta_{\text{\text{R2B}},il}^{\text{A}})$, $u_{\text{\text{R2B}},il}^{\text{D}} = \frac{2d_{\text{RIS}_i}}{\lambda} \cos(\vartheta_{\text{\text{R2B}},il}^{\text{D}}) \sin(\varphi_{\text{\text{R2B}},il}^{\text{D}})$ and $v_{\text{\text{R2B}},il}^{\text{D}} = \frac{2d_{\text{RIS}_i}}{\lambda} \sin(\vartheta_{\text{\text{R2B}},il}^{D})$, where $\vartheta_{\text{\text{R2B}},il}^{\text{A}}$ is the AoA at the $i$-th RIS for the $l$-th path, and $\vartheta_{\text{\text{R2B}},il}^{\text{D}}$ and $\varphi_{\text{\text{R2B}},il}^{\text{D}}$ are the elevation and azimuth AoDs from the BS towards the $i$-th RIS, respectively. $d_{\text{BS}}$ and $d_{\text{RIS}_i}$ are the BS antenna spacing and the $i$-th RIS element spacing, respectively, and $\lambda$ denotes the carrier wavelength.
	
	In general, RISs are deployed in the vicinity of UEs to significantly improve the quality of ISAC. Therefore, the LoS path from the $k$-th UE (\emph{treat the fixed site as the $0$-th UE} and $k \in \{0,1,...,K\}$) to the $i$-th RIS is given by
	\begin{equation}\label{hb}
		\mathbf{h}_{\text{b},ik} =\sqrt{N_i}  \alpha_{\text{b},ik} \mathbf{b}_{N_i}(u_{\text{\text{U2R}},ik}^{\text{A}}, v_{\text{\text{U2R}},ik}^{\text{A}}),
	\end{equation}
	where $\alpha_{\text{b},ik}$ represents the complex gain consisting of path loss for the path between the $i$-th RIS and $k$-th UE, $u_{\text{\text{U2R}},ik}^{\text{A}} = \frac{2d_{\text{RIS}_i}}{\lambda} \cos(\vartheta_{\text{\text{U2R}},ik}^{\text{A}}) \sin(\varphi_{\text{\text{U2R}},ik}^{\text{A}})$, $v_{\text{\text{U2R}},ik}^{\text{A}} = \frac{2d_{\text{RIS}_i}}{\lambda} \sin(\theta_{\text{\text{U2R}},ik}^{\text{A}})$, $\vartheta_{\text{\text{U2R}},ik}^{\text{A}}$ and $\varphi_{\text{\text{U2R}},ik}^{\text{A}}$ denotes elevation and azimuth AoAs at the $i$-th RIS from the $k$-th UE, respectively.
	
	The highly directional nature of propagation in mmWave channels prompts the spatial sparsity, which can be suitably leveraged to estimate the channel angles\cite{spatial}. In particular, we apply the basis expansion model on $\mathbf{H}_{\text{r},i}$ and $\mathbf{h}_{\text{b},ik}$ by discretizing the angle domain using a set of grids on the effective AoAs/AoDs to estimate $\{u_{\text{\text{R2B}},il}^{\text{A}}, \ u_{\text{\text{R2B}},il}^{\text{D}}, \ v_{\text{\text{R2B}},il}^{\text{D}} \}_{i,l=1,1}^{2,L_i}$ for $\mathbf{H}_{\text{r},i}$ and $\{u_{\text{\text{U2R}},ik}^{\text{A}}, v_{\text{\text{U2R}},ik}^{\text{A}}\}_{i,k=1,1}^{2,K}$ for $\mathbf{h}_{\text{b},ik}$. Then, Eq.~(\ref{Hr}) and Eq.~(\ref{hb}) can be rewritten as
	\begin{equation}\label{Hr_sparse}
		\mathbf{H}_{\text{r},i} = \sqrt{\frac{MN_i}{L_i}} \mathbf{A}_{\text{\text{R2B}},i}(\mathbf{w}_{\text{\text{R2B}},i}^{\text{A}}) \boldsymbol{\Omega}_{i} \mathbf{B}_{\text{\text{R2B}},i}^H(\mathbf{w}_{\text{\text{R2B}},i}^{\text{D}}, \mathbf{g}_{\text{\text{R2B}},i}^{\text{D}}),
	\end{equation}
	\begin{equation}\label{hb_sparse}
		\mathbf{h}_{\text{b},ik} = \sqrt{N_i} \mathbf{B}_{\text{\text{U2R}},i}(\mathbf{w}_{\text{\text{U2R}},i}^{\text{A}}, \mathbf{g}_{\text{\text{U2R}},i}^{\text{A}}) \boldsymbol{\psi}_{i,k},
	\end{equation}
	where $\mathbf{A}_{\text{\text{R2B}},i}(\mathbf{w}_{\text{\text{R2B}},i}^{\text{A}}) = [\mathbf{a}_M(w_{\text{\text{R2B}},i1}^{\text{A}}),\ \mathbf{a}_M(w_{\text{\text{R2B}},i2}^{\text{A}}), ..., \\ \mathbf{a}_M(w_{\text{\text{R2B}},i\text{G}_1}^{\text{A}})] \in \mathbb{C}^{M \times \text{G}_1}$, $\mathbf{B}_{\text{\text{R2B}},i}(\mathbf{w}_{\text{\text{R2B}},i}^{\text{D}}, \mathbf{g}_{\text{\text{R2B}},i}^{\text{D}}) =[\mathbf{b}(w_{\text{\text{R2B}},i1}^{\text{D}}, g_{\text{\text{R2B}},i1}^{\text{D}}),...,\mathbf{b}(w_{\text{\text{R2B}},i\text{G}_2}^{\text{D}}, g_{\text{\text{R2B}},i1}^{\text{D}}),\mathbf{b}(w_{\text{\text{R2B}},i1}^{\text{D}}, g_{\text{\text{R2B}},i2}^{\text{D}})\\,...,\mathbf{b}(w_{\text{\text{R2B}},i\text{G}_2}^{\text{D}}, g_{\text{\text{R2B}},i\text{G}_3}^{\text{D}})] \in \mathbb{C}^{N_i \times \text{G}_2 \text{G}_3}$ and $\mathbf{B}_{\text{\text{U2R}},i}(\mathbf{w}_{\text{\text{U2R}},i}^{\text{A}},\\ \mathbf{g}_{\text{\text{U2R}},i}^{\text{A}}) = [\mathbf{b}(w_{\text{\text{U2R}},i1}^{\text{A}}, g_{\text{\text{U2R}},i1}^{\text{A}}),...,\mathbf{b}(w_{\text{\text{U2R}},i\text{G}_4}^{\text{A}}, g_{\text{\text{U2R}},i1}^{\text{A}}),\\ \mathbf{b}(w_{\text{\text{U2R}},i1}, g_{\text{\text{U2R}},i2}^{\text{A}}),...,\mathbf{b}(w_{\text{\text{U2R}},i\text{G}_4}^{\text{A}}, g_{\text{\text{U2R}},i\text{G}_5}^{\text{A}})]  \in  \mathbb{C}^{N_i \times \text{G}_4 \text{G}_5}$ are dictionary matrices with $\mathbf{w}_{\text{\text{R2B}},i}^{\text{A}}$, $\mathbf{w}_{\text{\text{R2B}},i}^{\text{D}}$, $\mathbf{g}_{\text{\text{R2B}},i}^{\text{D}}$, $\mathbf{w}_{\text{\text{U2R}},i}^{\text{A}}$ and $\mathbf{g}_{\text{\text{U2R}},i}^{\text{A}}$ being quantized grids for effective AoAs/AoDs: $\{u_{\text{\text{R2B}},il}^{\text{A}}\}_{l=1}^{L_i}$, $\{u_{\text{\text{R2B}},il}^{\text{D}}\}_{l=1}^{L_i}$, $\{ v_{\text{\text{R2B}},il}^{\text{D}} \}_{l=1}^{L_i}$, $\{u_{\text{\text{U2R}},ik}^{\text{A}}\}_{k=1}^{K}$ and $\{v_{\text{\text{U2R}},ik}^{\text{A}}\}_{k=1}^{K}$, respectively, for $\forall i$.
	
	\begin{remark}
		\emph{Here it is actually easy to determine the $1$-st path between the BS and RISs, i.e., $[u_{\text{\text{R2B}},i1}^{\text{A}}, u_{\text{\text{R2B}},i1}^{\text{D}}, v_{\text{\text{R2B}},i1}^{\text{D}}]_{i=1}^{2}$, which corresponds to effective AoAs/AoDs of LoS paths, and can be obtained from the relative positions. Thus, the first elements of $\mathbf{w}_{\text{R2B},i}^{\text{A}}$, $\mathbf{w}_{\text{\text{R2B}},i}^{\text{D}}$ and $\mathbf{g}_{\text{\text{R2B}},i}^{\text{D}}$ are set to $u_{\text{\text{R2B}},i1}^{\text{A}}$, $u_{\text{\text{R2B}},i1}^{\text{D}}$ and $v_{\text{\text{R2B}},i1}^{\text{D}}$, respectively, while we only need to use uniform quantization to divide the remaining $[u_{\text{\text{R2B}},il}^{\text{A}}, u_{\text{\text{R2B}},il}^{\text{D}}, v_{\text{\text{R2B}},il}^{\text{D}}]_{i,l=1,2}^{2,L_i}$ evenly into $\text{G}_1-1$, $\text{G}_2-1$ and $\text{G}_3-1$ portions. Then, effective AoAs from each UE, i.e., $[u_{\text{\text{U2R}},ik}^{\text{A}}, v_{\text{\text{U2R}},ik}^{\text{A}}]_{i=1,k}^{2,K}$, are uniformly quantized into $\text{G}_4$ and $\text{G}_5$ portions, where grids are included into $\mathbf{w}_{\text{\text{U2R}},i}^{\text{A}}$ and $\mathbf{g}_{\text{\text{U2R}},i}^{\text{A}}$.}
	\end{remark}
	
	If actual AoAs/AoDs $\{u_{\text{\text{R2B}},il}^{\text{A}}, \  u_{\text{\text{R2B}},il}^{\text{D}}, \ v_{\text{\text{R2B}},il}^{\text{D}} \}_{i,l=2,2}^{2,L_i}$ and $\{u_{\text{\text{U2R}},ik}^{\text{A}}, \ v_{\text{\text{U2R}},ik}^{\text{A}}\}_{i,k=1}^{2,K}$ lie on the selected grids, there will be $L_i$ and one non-zero elements in $\boldsymbol{\Omega}_i$ and $\boldsymbol{\psi}_{i,k}$, respectively. However, this is usually untrue, and mismatch model errors will incur incorrect angle estimates. Hence, off-grid methods are needed to fine-tune the grids delicately.
	
	\subsection{Signal Model} 
	The transmitted signal for ISAC periods is designed to achieve the dual functionalities of communication and localization, which is proposed to bear data symbols and sensing pilots simultaneously. Through this scheme, we aim to explore advanced signal processing techniques to integrate channel angle sensing and localization capabilities into the existing multi-UE communication process. Specifically, received signals at the BS can be expressed in separate frequency bands for the fixed site and multiple UEs, respectively, as follows.
	
	\subsubsection{Received signals from the fixed site}
	Focusing on the $T_1$ ISAC period, the fixed site transmits $x_{0,t}$ at the $t$-th time slot with $|x_{0,t}|^2 \le P_{0}$. Then, the received signal at the BS is
	\begin{equation}\label{y0t}
		\mathbf{y}_{0,t} = \sum\limits_{i=1}^2 \mathbf{H}_{\text{r},i} \text{diag}(\boldsymbol{\theta}_{i,t}) \mathbf{h}_{\text{b},i0} x_{0,t} + \mathbf{n}_{0,t},
	\end{equation}
	where $\boldsymbol{\theta}_{i,t}=[e^{j\theta_{t,1}},...,e^{j \theta_{t,N_i}}]^T$ denotes the $i$-th RIS phase shifts at the $t$-th time slot, and $\mathbf{n}_{0,t}$ is additive white Gaussian noise follows $ \mathcal{CN}(\mathbf{0},N_0 \mathbf{I})$. The transmitted symbol can be expressed as $x_{0,t}=\sqrt{\xi_0}x_{\text{d},0,t}+\sqrt{1-\xi_0}x_{\text{p},0,t}$, where $x_{\text{d},0,t}$ and $x_{\text{p},0,t}$ denote the data and sensing pilot symbol, respectively, and $\xi_0$ represents the power proportion factor of data with $0<\xi_0<1$.
	
	\subsubsection{Received signals from multiple UEs}
	We allocate $N_\text{s}$ frequency bands for multi-UE ISAC period, where the $k$-th UE transmits $x_{k,t,n_\text{s}}$ at the $t$-th time slot for the $n_\text{s}$-th frequency band. Then, the received signal at the BS can be expressed as
	\begin{equation}\label{yUt}
		\mathbf{y}_{\text{U},t,n_\text{s}} = \sum\limits_{k=1}^K \sum\limits_{i=1}^2 \mathbf{H}_{\text{r},i} \text{diag}(\boldsymbol{\theta}_{i,t}) \mathbf{h}_{\text{b},ik} x_{k,t,n_\text{s}} +\mathbf{n}_{\text{U},t,n_\text{s}},
	\end{equation}
	for $t \in \{1,...,T_2\}$ and $n_\text{s} \in \{1,...,N_\text{s}\}$, where $\mathbf{n}_{\text{U},t,n_\text{s}}$ follows $ \mathcal{CN}(\mathbf{0},N_0 \mathbf{I})$. The transmitted symbol can be expressed as $x_{k,t,n_\text{s}}=\sqrt{\xi_k}x_{\text{d},k,t,n_\text{s}}+\sqrt{1-\xi_k}x_{\text{p},k,t,n_\text{s}}$, where $x_{\text{d},k,t,n_\text{s}}$ and $x_{\text{p},k,t,n_\text{s}}$ denote the data and sensing pilot symbol, respectively, and $\xi_k$ represents the power proportion factor of data with $0<\xi_k<1$.
	
	In the following, we develop an efficient Bayesian learning-based method to obtain data symbols and channel angle information of BS-RIS links by (\ref{y0t}). Then, we propose the SCMA-UAMP-SBL method to achieve simultaneous communication and localization for multiple UEs by (\ref{yUt}).

	\section{ISAC Period for the Fixed site}\label{sec_fix}
	After collecting received measurements along $T_1$ time slots from (\ref{y0t}), we can obtain $\mathbf{Y}_0 = [\mathbf{y}_{0,1},...,\mathbf{y}_{0,T_1}] \in \mathbb{C}^{M \times T_1}$ as
	\begin{equation}\label{Y0}
		\mathbf{Y}_{0} = \sum\limits_{i=1}^2 \mathbf{H}_{\text{r},i} \text{diag}(\mathbf{h}_{\text{b},i0}) \boldsymbol{\Theta}_i \mathbf{X}_0 + \mathbf{N}_0,
	\end{equation}
	where $\boldsymbol{\Theta}_{i} = [\boldsymbol{\theta}_{i,1},...,\boldsymbol{\theta}_{i,T_1}]$, $\mathbf{X}_0 = \text{diag}(x_{0,1},...,x_{0,T_1})$ and $\mathbf{N}_0 = [\mathbf{n}_{0,1},...,\mathbf{n}_{0,T_1}]$. By introducing the mmWave sparse model of $\mathbf{H}_{\text{r},i}$ as in Eq.~(\ref{Hr_sparse}), the received signal yields
	\begin{equation}\label{Ys}
		\mathbf{Y}_0 = \sum\limits_{i=1}^2 \sqrt{\frac{M N_i}{L_i}} \mathbf{A}_{\text{\text{R2B}},i} \boldsymbol{\Omega}_{i} \mathbf{B}_{\text{\text{R2B}},i}^H \text{diag}(\mathbf{h}_{\text{b},i0}) \boldsymbol{\Theta}_i \mathbf{X}_0 + \mathbf{N}_0,
	\end{equation}
	where $\mathbf{A}_{\text{\text{R2B}},i}(\mathbf{w}_{\text{\text{R2B}},i}^{\text{A}})$ and $\mathbf{B}_{\text{\text{R2B}},i}^H(\mathbf{w}_{\text{\text{R2B}},i}^{\text{D}}, \mathbf{g}_{\text{\text{R2B}},i}^{\text{D}})$ are abbreviated to $\mathbf{A}_{\text{\text{R2B}},i}$ and $\mathbf{B}_{\text{\text{R2B}},i}^H$, respectively. Considering the vectorization $\mathbf{y}_0 = \text{vec}(\mathbf{Y}_0)$ in Eq.~(\ref{Ys}), and employing the property $\text{vec}(\mathbf{A} \mathbf{B}\mathbf{C})=(\mathbf{C}^T \otimes \mathbf{A})\text{vec}(\mathbf{B}$), where $\otimes$ is the Kronecker product, we have
	\begin{equation}\label{y0}
		\begin{aligned}
			\mathbf{y}_0 &= \sum\limits_{i=1}^2 \sqrt{\frac{M N_i}{L_i}} \left[ \left( \mathbf{B}_{\text{\text{R2B}},i}^H \text{diag}(\mathbf{h}_{\text{b},i0}) \boldsymbol{\Theta}_i \mathbf{X}_0 \right)^T \otimes \mathbf{A}_{\text{\text{R2B}},i} \right] \\
			& \qquad \cdot \text{vec}(\boldsymbol{\Omega}_i) + \mathbf{n}_0\ = \mathbf{Z}_{\text{\text{R2B}}}(\mathbf{X}_0,\boldsymbol{\nu}) \boldsymbol{\omega} + \mathbf{n}_0,
		\end{aligned}
	\end{equation}
	where $\mathbf{Z}_{\text{\text{R2B}}}(\mathbf{X}_0, \boldsymbol{\nu}) = [\mathbf{Z}_{\text{\text{R2B}},1}(\mathbf{X}_0,\boldsymbol{\nu}_1), \ \mathbf{Z}_{\text{\text{R2B}},2}(\mathbf{X}_0, \boldsymbol{\nu}_2)]$,
	$\mathbf{Z}_{\text{\text{R2B}},i}(\mathbf{X}_0,\boldsymbol{\nu}_i) = \sqrt{\frac{M N_i}{L_i}} \left( \mathbf{B}_{\text{\text{R2B}},i}^H \text{diag}(\mathbf{h}_{\text{b},i0}) \boldsymbol{\Theta}_i \mathbf{X}_0 \right)^T \otimes \mathbf{A}_{\text{\text{R2B}},i}$, $\boldsymbol{\nu}_i = [\mathbf{w}_{\text{\text{R2B}},i}^{\text{A}}, \mathbf{w}_{\text{\text{R2B}},i}^{\text{D}}, \mathbf{g}_{\text{\text{R2B}},i}^{\text{D}} ]$, $\boldsymbol{\omega} = [\boldsymbol{\omega}_1 ; \ \boldsymbol{\omega}_2]$, and $\boldsymbol{\omega}_i = \text{vec}(\boldsymbol{\Omega}_i)$. This indicates that precise symbols $\mathbf{X}_0$ and refined grids $\boldsymbol{\nu}$ are needed to yield accurate $\boldsymbol{\omega}$, which in turn facilitates the estimates of $\mathbf{X}_{\text{d},0}$ and $\boldsymbol{\nu}$. Thus, an iterative framework is developed for updating $\mathbf{X}_{\text{d},0}$, $\boldsymbol{\nu}$ and $\boldsymbol{\omega}$. In the following, we first determine initial data estimates. Then, a structure-aware Bayesian learning-based method is proposed to obtain effective AoAs/AoDs and data estimates efficiently.  
	
	\subsection{Initial Estimates by the TR-based LS Method}
	The first step is to recover the coarse estimate $\boldsymbol{\omega}$ from sensing pilots $\mathbf{X}_{\text{p},0}$. We can divide the sensing matrix in (\ref{y0}) into two parts, i.e., $\mathbf{Z}_{\text{\text{R2B}}}(\mathbf{X}_0,\boldsymbol{\nu})=\mathbf{Z}_{\text{\text{R2B}}}(\sqrt{\xi_0}\mathbf{X}_{\text{d},0})+\mathbf{Z}_{\text{\text{R2B}}}(\sqrt{1-\xi_0}\mathbf{X}_{\text{p},0})$ ignoring $\boldsymbol{\nu}$, and formulate the least square estimate (LSE) with Tikhonov regularization (TR) \cite{TR} as
	\begin{equation}\label{tr2}
		\begin{aligned}
			\arg \min_{\boldsymbol{\omega}}  \quad & \|\mathbf{y}_0 - \mathbf{Z}_{\text{\text{R2B}}}(\sqrt{1-\xi_0} \mathbf{X}_{\text{p},0})\boldsymbol{\omega}\|^2 +\\
			&\qquad \varrho^2 \|\mathbf{Z}_{\text{\text{R2B}}}(\sqrt{\xi_0} \mathbf{X}_{\text{d},0}) \boldsymbol{\omega}\|^2,
		\end{aligned}
	\end{equation}
	where $\varrho$ denotes the data uncertainty. The introduction of TR solves the ill-posed problem in LSE. The second term of (\ref{tr2}) can be calculated as  $\varrho^2 \boldsymbol{\omega}^H \boldsymbol{\Upsilon} \boldsymbol{\omega}$ in the closed form, where $\boldsymbol{\Upsilon} 
	=
	\begin{bmatrix}
		\boldsymbol{\Upsilon}_{1,1}^S	&  \boldsymbol{\Upsilon}_{1,2}^S \\
		\boldsymbol{\Upsilon}_{2,1}^S	& \boldsymbol{\Upsilon}_{2,2}^S
	\end{bmatrix}$
	and $\boldsymbol{\Upsilon}_{i,j}^S = \xi_0 \sqrt{\frac{M^2 N_i N_j}{L_i L_j}} \big[ ( \mathbf{B}_{\text{\text{R2B}},i}^H  \text{diag}(\mathbf{h}_{\text{b},i0}) \boldsymbol{\Theta}_i )^* \big( \mathbf{B}_{\text{\text{R2B}},j}^H \text{diag}(\mathbf{h}_{\text{b},j0})\\ \boldsymbol{\Theta}_j \big)^T \big]\otimes \left(  \mathbf{A}_{\text{\text{R2B}},i}^H \mathbf{A}_{\text{\text{R2B}},j} \right)$.
	
	Then, the initialization solution to $\boldsymbol{\omega}$ is obtained by
	\begin{equation}\label{w0}
		\begin{aligned}
			\boldsymbol{\omega}^{(0)} &= \left[ \mathbf{Z}_{\text{\text{R2B}}}^H(\sqrt{1-\xi_0} \mathbf{X}_{\text{p},0}) \mathbf{Z}_{\text{\text{R2B}}}(\sqrt{1-\xi_0} \mathbf{X}_{\text{p},0}) + \varrho^2  \boldsymbol{\Upsilon} \right]^{-1} \\
			& \qquad\cdot \mathbf{Z}_{\text{\text{R2B}}}(\sqrt{1-\xi_0} \mathbf{X}_{\text{p},0})^H \mathbf{y}_0.
		\end{aligned}
	\end{equation}
	Based on this estimate, initial data observation estimates can refer to $\mathbf{\hat{Y}}_{\text{d},0} = \mathbf{Y}_0 - \mathbf{H}^{(0)}_{0,\text{eff}} \sqrt{1-\xi_0} \mathbf{X}_{\text{p},0}$, where the effective channel estimate is expressed as $\mathbf{H}^{(0)}_{0,\text{eff}} = \sum_{i=1}^2  \sqrt{\frac{M N_i}{L_i}} \mathbf{A}_{\text{\text{R2B}},i} \text{vec}^{-1}(\boldsymbol{\omega}_i^{(0)}) \mathbf{B}_{\text{\text{R2B}},i}^H \text{diag}(\mathbf{h}_{i,0}) \boldsymbol{\Theta}_i$, and $\text{vec}^{-1}(\cdot)$ represents the inverse vectorization process. Hence, the linear minimum mean square error (LMMSE) solution of $\mathbf{X}_{\text{d},0}$ is given by
	\begin{equation}\label{X0}
		\begin{aligned}
			\mathbf{X}_{\text{d},0}^{(0)} = \frac{1}{\sqrt{\xi_0}} &\left( \left( \mathbf{H}^{(0)}_{0,\text{eff}} \right)^H  \mathbf{H}^{(0)}_{0,\text{eff}} + \frac{N_0}{P_0} \mathbf{I} \right)^{-1} \left( \mathbf{H}^{(0)}_{0,\text{eff}} \right)^H \\
			& \cdot \left( \mathbf{Y}_0 - \sqrt{1-\xi_0} \mathbf{H}^{(0)}_{0,\text{eff}} \mathbf{X}_{\text{p},0}  \right).
		\end{aligned}
	\end{equation}
	
	\subsection{Structure-aware Bayesian learning framework}
	In this section, we exploit the sparse structure of $\boldsymbol{\omega}$ and superimposed structure of $\mathbf{X}_0$ to update $\boldsymbol{\omega}$, $\boldsymbol{\nu}$ and $\mathbf{X}_{\text{d},0}$ by using the Bayesian learning in expectation maximization (EM) procedures. Specifically, we first set \emph{a priori} distribution for $\boldsymbol{\omega}$ to promote the sparsity. Then, the log-likelihood function in terms of $\boldsymbol{\omega}$, $\boldsymbol{\nu}$ and $\mathbf{X}_{\text{d},0}$ in Bayesian framework is formulated, which will be maximized in an iterative manner.
	
	Inspired by SBL on channel angular domain \cite{SBL,SBL2,SBL3}, a fictitious sparsity promoting zero-mean Gaussian prior with unknown covariance is assumed on $\boldsymbol{\omega}$ as
	\begin{equation}\label{om_pri}
		p(\boldsymbol{\omega};\boldsymbol{\gamma}) = \mathcal{CN}(\boldsymbol{\omega} | \mathbf{0}, \text{diag}(\boldsymbol{\gamma}^{-1})),
	\end{equation}
	where $\boldsymbol{\gamma}=[\gamma_1,...,\gamma_{2\text{G}_1\text{G}_2\text{G}_3}]$, and $\gamma_g^{-1}$ denotes the variance corresponding to $\omega_g$, for $\forall g \in \{1,...,2\text{G}_1\text{G}_2\text{G}_3\}$. Based on the prior assignment in Eq.~(\ref{om_pri}), one can observe that as $\gamma_g^{-1} \rightarrow 0$, the associated $\omega_g \rightarrow 0$. Thus, the estimation of $\omega_g$ is reduced to the estimation of the hyperparameter vector $\boldsymbol{\gamma}$. To further encourage sparsity on the prior assignment on $\boldsymbol{\omega}$, we model ${\gamma}_g$ as independent Gamma distributions, i.e., $p(\boldsymbol{\gamma}) = \prod\limits_{g=1}^{2\text{G}_1\text{G}_2\text{G}_3} \Gamma(\gamma_g;1+a_{\gamma},b_{\gamma})$, where we set $a_{\boldsymbol{\gamma}},b_{\boldsymbol{\gamma}} \to 0$ due to the heavy tails and sharp peak at zero.
	
\begin{figure}[ht]
		\centering
		\includegraphics[scale=0.2]{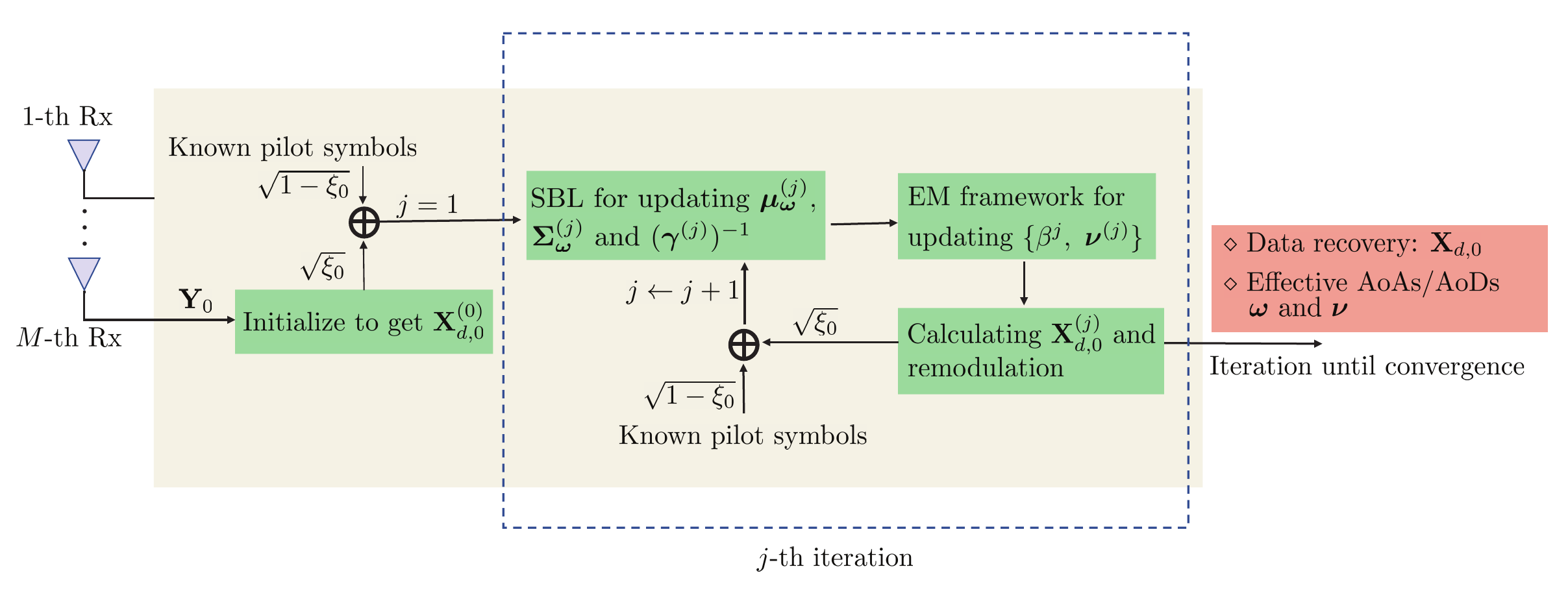}\vspace{-2pt}
		\caption{Flow chart for ISAC signal processing.}
		\label{fix_site_block}
	\end{figure}

	Then, in the EM procedure, we give the complete information set as $\{\mathbf{y}_0, \boldsymbol{\omega}\}$, and the parameter set as $\boldsymbol{\chi} = [\mathbf{X}_{\text{d},0}, \boldsymbol{\gamma}, \boldsymbol{\nu}, \beta ]$, where $\beta$ is modeled to handle the uncertainty of unknown noise variance in practice. Then the EM-based Bayesian learning to estimate $\boldsymbol{\chi}$ is developed as in Fig.~\ref{fix_site_block} and divided into two steps as follows.
	
	\subsubsection{E-Step}
	We first evaluate the log-likelihook function $\mathcal{L}(\boldsymbol{\chi}|\boldsymbol{\chi}^{(j-1)})$ as the cost function, where the superscript $(j-1)$ represents variables obtained in the $(j-1)$-th iteration, expressed as
	\begin{equation}\label{li}
		\begin{aligned}
			&\mathcal{L}(\boldsymbol{\chi}|\boldsymbol{\chi}^{(j-1)})  = \mathbb{E}_{\boldsymbol{\omega} | \mathbf{y}_0 ; \boldsymbol{\chi}^{(j-1)}}\big\{\ln \big[ p(\mathbf{y}_0, \boldsymbol{\omega}; \boldsymbol{\chi}) \big] \big\} \\
			& = \mathbb{E}_{\boldsymbol{\omega} | \mathbf{y}_0 ; \boldsymbol{\chi}^{(j-1)}} \bigg\{\ln \big[p(\mathbf{y}_0|  \boldsymbol{\omega};\mathbf{X}_{\text{d},0}, \boldsymbol{\nu}, \beta ) \big]  + \ln  \big[p(\boldsymbol{\omega}; \boldsymbol{\gamma})  \big] \bigg\},
		\end{aligned}
	\end{equation}
	where the channel posterior density can be expressed as $p(\boldsymbol{\omega} | \mathbf{y}_0; \boldsymbol{\chi}^{(j-1)}) \sim \mathcal{CN}(\boldsymbol{\mu}_{\boldsymbol{\omega}}^{(j-1)},\boldsymbol{\Sigma}_{\boldsymbol{\omega}}^{(j-1)})$. The closed-form expressions via \cite{SBL} are given as $\boldsymbol{\Sigma}_{\boldsymbol{\omega}}^{(j-1)}=\bigg(\beta^{(j-1)}  \mathbf{Z}_{\text{\text{R2B}}}^H \\(\mathbf{X}_0^{(j-1)},\boldsymbol{\nu}^{(j-1)}) \mathbf{Z}_{\text{\text{R2B}}}(\mathbf{X}_0^{(j-1)},\boldsymbol{\nu}^{(j-1)})+ \text{diag}(\boldsymbol{\gamma}^{(j-1)})\bigg)^{-1}$ and
	$\boldsymbol{\mu}_{\boldsymbol{\omega}}^{(j-1)} = \beta^{(j-1)} \boldsymbol{\Sigma}_{\boldsymbol{\omega}}^{(j-1)} \mathbf{Z}_{\text{\text{R2B}}}^H(\mathbf{X}_0^{(j-1)},\boldsymbol{\nu}^{(j-1)}) \mathbf{y}_0$.
	
	\begin{remark}
		\emph{Theoretically, there are only limited non-zero elements in $\boldsymbol{\mu}_{\boldsymbol{\omega}}^{(j-1)}$ due to the mmWave scattering characteristics. However, in practice, $\boldsymbol{\mu}_{\boldsymbol{\omega}}^{(j-1)}$ may contain a large number of very small values, but not zero, which can impair the performance of angle refinements and data detection. To address this degradation, a method is proposed to select a few elements with large values, such as those greater than $\delta_{\omega}|\boldsymbol{\mu}_{\boldsymbol{\omega}}^{(j-1)}|$ with $0<\delta_{\omega}<1$, and set the others to zero. The impact of these zero settings can be compensated by the proposed LSE method in \cite{iter_huang}.	}
	\end{remark}

	Notice that the first term of (\ref{li}) depends on $\mathbf{X}_{\text{d},0}$, $\boldsymbol{\nu}$ and $\beta$, and the second term only depends on $\boldsymbol{\gamma}$. Hence, the following \emph{M-step} can be reduced to two independent maximizations of (\ref{li}) with respect to (w.r.t) $\boldsymbol{\gamma}$ and $\{\mathbf{X}_{\text{d},0}, \boldsymbol{\nu}, \beta\}$, respectively. 
	
	\subsubsection{M-Step \uppercase\expandafter{\romannumeral1}}
	In response to the maximization of the second term in (\ref{li}), the hyperparameters yield
	\begin{equation}\label{gamma}
		\gamma_g^{(j)} = \frac{a_{{\gamma}}+1}{b_{{\gamma}} + \boldsymbol{\Sigma}_{\boldsymbol{\omega}}^{(j-1)}(g,g)+|\boldsymbol{\mu}_{\boldsymbol{\omega}}^{(j-1)}(g)|^2},
	\end{equation}
	where $\boldsymbol{\Sigma}_{\boldsymbol{\omega}}^{(j-1)}(g,g)$ is the $(g,g)$-th element of $\boldsymbol{\Sigma}_{\boldsymbol{\omega}}^{(j-1)}$, and $\boldsymbol{\mu}_{\boldsymbol{\omega}}^{(j-1)}(g)$ is the $g$-th element of $\boldsymbol{\mu}_{\boldsymbol{\omega}}^{(j-1)}$.
	
	\subsubsection{M-Step \uppercase\expandafter{\romannumeral2}}
	In this step, we update each parameter in $\{\mathbf{X}_{\text{d},0}, \boldsymbol{\nu}, \beta\}$ at a time while fixing the others, which can yield preferable performance in this mutual coupling case. 
	
	Note that we only need to update grids $\boldsymbol{\nu}_2 = [w_{\text{\text{R2B}},il}^{\text{A}}, w_{\text{\text{R2B}},il}^{\text{D}}, g_{\text{\text{R2B}},il}^{\text{D}}]_{i,l=1,2}^{2,L_i}$ regardless of LoS paths as in \emph{Remark 1}. Then, the update of $\beta^{(j)}$ and $\boldsymbol{\nu}_2^{(j)}$ can be obtained, respectively,
	\begin{equation}\label{beta^j}
		\begin{aligned}
			\beta^{(j)} &= (MT_1)/ \bigg{(} \|\mathbf{y}_0 - \mathbf{Z}_{\text{\text{R2B}}}(\mathbf{X}_0^{(j-1)}, \boldsymbol{\nu}^{(j-1)}) \boldsymbol{\mu}_{\boldsymbol{\omega}}^{(j-1)}\|^2+\\
			&\ \text{Tr} \big{(} \mathbf{Z}_{\text{\text{R2B}}}(\mathbf{X}_0^{(j-1)}, \boldsymbol{\nu}^{(j-1)}) \boldsymbol{\Sigma}_{\boldsymbol{\omega}}^{(j-1)} \mathbf{Z}_{\text{\text{R2B}}}^H(\mathbf{X}_0^{(j-1)}, \boldsymbol{\nu}^{(j-1)}) \big{)}  \bigg{)},
		\end{aligned}
	\end{equation}
	\begin{equation}\label{nu^j}
		\boldsymbol{\nu}_2^{{(j)}} = \boldsymbol{\nu}_2^{(j-1)} + \epsilon_1^{(j)}  \mathcal{F}_1(\boldsymbol{\nu}_2^{(j-1)}),
	\end{equation}
	where $\epsilon_1^{(j)}$ is a dynamically chosen step-size sequence defined in (\ref{epsilon1}), $\mathcal{F}_1(\boldsymbol{\nu}_2^{(j-1)})$ is the closed-form derivative of (\ref{nu}), and the detailed derivation process can be found in Appendix \ref{beta_nu}. Then, we derive the update for $\mathbf{X}_{\text{d},0}$ with $\boldsymbol{\nu}^{(j)}$ and $\beta^{(j)}$ as in \emph{Lemma~\ref{lem}}. 
	\begin{lemma}\label{lem}
		By maximizing the first term of (\ref{li}), the estimated data yields
		\begin{equation}\label{xd}
			\begin{aligned}
				\mathbf{X}_{\text{d},0}^{(j)} =& \frac{1}{\sqrt{\xi_0}}\left(\left(\boldsymbol{\mu}_{H\text{eff}}^{(j-1)}\right)^H \boldsymbol{\mu}_{H\text{eff}}^{(j-1)} + \boldsymbol{\Sigma}_{H\text{eff}}^{(j-1)}\right)^{-1}\bigg(\left(\boldsymbol{\mu}_{H\text{eff}}^{(j-1)}\right)^H \cdot \\
				&\mathbf{Y}_0 -\left(\left(\boldsymbol{\mu}_{H\text{eff}}^{(j-1)}\right)^H \boldsymbol{\mu}_{H\text{eff}}^{(j-1)} + \boldsymbol{\Sigma}_{H\text{eff}}^{(j-1)} \right) \sqrt{1-\xi_0} \mathbf{X}_{\text{p},0} \bigg),
			\end{aligned}
		\end{equation}
		where $\boldsymbol{\mu}_{H\text{eff}}^{(j-1)}$ and $\boldsymbol{\Sigma}_{H\text{eff}}^{(j-1)}$ are given in Eq.~(\ref{mu_H}) and Eq.~(\ref{sigma_H}), respectively.
	\end{lemma}
	\proof
	See Appendix \ref{X_d,0}.
	\endproof
	
	\begin{remark}
		\emph{To further improve the accuracy of data estimation after each iteration, the good anti-noise performance of digital modulations is utilized. Specifically, the data estimate is demodulated into information bits and then modulated into symbols. Therefore, the data estimate for the $j$-th iteration is replaced by $\text{Mod}(\text{Demod}(\mathbf{X}_{\text{d},0}^{(j)}))$.}
	\end{remark}

	Repeat the above process until the difference between $\mathcal{L}(\boldsymbol{\chi}|\boldsymbol{\chi}^{(j)})$ and $\mathcal{L}(\boldsymbol{\chi}|\boldsymbol{\chi}^{(j-1)})$ is less than the threshold $\delta_{\chi}$ or the number of iterations is up to $j_{\max}$. Then, the maximum \emph{a posteriori} probability estimate yields $\boldsymbol{\hat{\omega}}=\boldsymbol{\mu}_{\boldsymbol{\omega}}^{(J)}$, where $J$ is the number of iterations at convergence. The estimates of refined grids and data symbols are $\boldsymbol{\nu}^{(J)}$ and $\mathbf{X}_{\text{d},0}^{(J)}$, respectively. The procedures of this structure-aware Bayesian learning algorithm during $T_1$ ISAC period are summarized in \emph{Algorithm~1}.
	\begin{algorithm}
		\caption{Structure-aware Bayesian Learning Algorithm for the ISAC Period}
		\begin{algorithmic}[1]		
			\STATE Initialize the coarse estimates $\boldsymbol{\omega}^{(0)}$ and $\mathbf{X}_{\text{d},0}^{(0)}$ by (\ref{w0}) and (\ref{X0}), respectively.
			\STATE Initialize $\boldsymbol{\gamma}^{(0)} = \mathbf{1}$, $a_{\gamma}=b_{\gamma}=10^{-4}$, $\beta^{(0)}=1$ and the iteration number $j=1$.
			\REPEAT
			\STATE Update channel angle parameters $\boldsymbol{\gamma}^{(j)}$, $\beta^{(j)}$ and $\boldsymbol{\nu}^{(j)}$ based on $\boldsymbol{\mu}_{\boldsymbol{\omega}}^{(j-1)}$ and $\boldsymbol{\Sigma}_{\boldsymbol{\omega}}^{(j-1)}$ by (\ref{gamma}), (\ref{beta^j}) and (\ref{nu^j}), respectively, in an iterative manner.
			\STATE Refine $\boldsymbol{\mu}_{\boldsymbol{\omega}}^{(j-1)}$ as in \emph{Remark 2}.
			\STATE Update data symbols $\mathbf{X}_{\text{d},0}^{(j)}$ by (\ref{xd}).
			\STATE Replace $\mathbf{X}_{\text{d},0}^{(j)}$ by  $\text{Mod} \left[\text{Demod}(\mathbf{X}_{\text{d},0}^{(j)}) \right]$.
			\STATE $j \gets j+1	$
			
			\UNTIL $ |\mathcal{L}(\boldsymbol{\chi}|\boldsymbol{\chi}^{(j)}) - \mathcal{L}(\boldsymbol{\chi}|\boldsymbol{\chi}^{(j-1)}) |^2  < \delta_{\chi} $ or $j>j_{\max} $ 
			
		\end{algorithmic}  
	\end{algorithm}

	\section{Multi-user location sensing and communication}
	In this section, we design a low-complexity algorithm for simultaneous communication and localization based on the superimposed symbol scheme in double-RIS aided ISAC systems. Specifically, the proposed algorithm employs UAMP-SBL to avoid matrix inversion operations for initial channel angle estimates of UEs. Then, estimates are refined iteratively through reduced-dimension matrix calculations. Moreover, the SCMA technology is utilized to effectively suppress interference between multiple UEs for accurate data detection, which further facilitates the accuracy of localization.
	
\begin{figure*}
		\centering
		\includegraphics[scale=0.32]{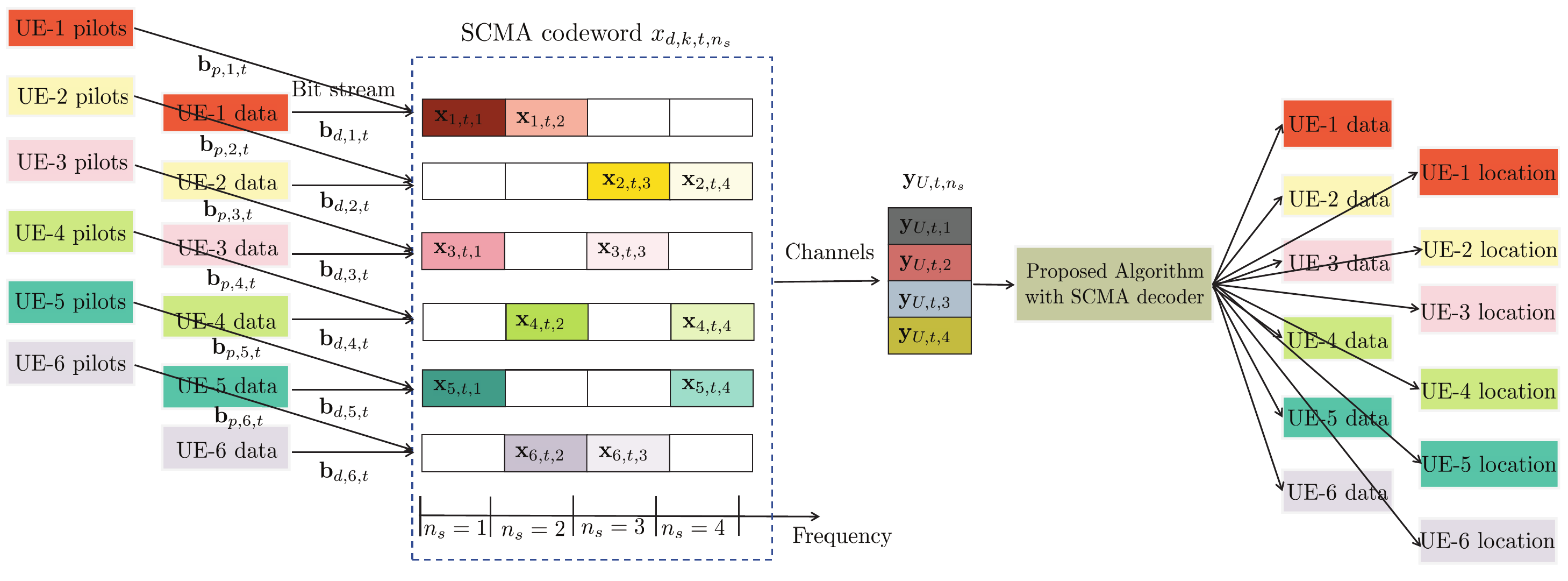}\vspace{-2pt}
		\caption{Flow chart for multi-UE SCMA encoding and decoding process for the case with $K=6$ and $N_\text{s}=4$, where the pilots and data of $K$ UEs will be encoded by SCMA codebook and transmitted simultaneously, then our algorithm is proposed to decode the communication data and obtain the location coordinates of $K$ UEs at the receiver side.}
		\label{SCMA_flow}
	\end{figure*}

	We first illustrate the block diagram for this uplink ISAC system in Fig.~\ref{SCMA_flow}. Specifically, the data bit stream for the $k$-th UE at the $t$-th time slot $\mathbf{b}_{\text{d},k,t}$ is first sent to the SCMA encoder, in which the codeword $\mathbf{x}_{\text{d},k,t}=[x_{\text{d},k,t,1},...,x_{\text{d},k,t,N_\text{s}}] \in \mathbb{C}^{1 \times N_\text{s}}$ is generated. This is equivalent to mapping elements by the SCMA codebook, i.e., $\mathbf{x}_{\text{d},k,t}=f(\mathbf{b}_{\text{d},k,t})$, $f : \ \mathbb{B}^{\log_2(N_{\text{c}})}\to \mathcal{X}$, where $N_{\text{c}}$ is the cardinality of the codebook. Then, $\mathbf{x}_{\text{d},k,t}$ is sparse such that only $d_v \ll N_\text{s}$ dimensions are used to transmit data while the remaining ones are set to be zeros. Thanks to the sparsity of SCMA codewords, the number of overlapped UEs in each subcarrier equals $d_c \ll K$. In this way, the effect of mitigating interference between UEs can be achieved.
	
	\begin{remark}
		\emph{It should be noted that in addition to inter-UE interference, mutual interference also exists between pilots and data through the superimposed symbol scheme. Actually, the pilots of the $k$-th UE can interfere with the data of all UEs due to imperfect channel estimates. This is a major challenge of using superimposed symbols for multi-UE scenarios. To address this challenge, an iterative framework approach is designed to mitigate this mutual interference. Additionally, to preserve the elaborately constructed sparsity of SCMA codewords, the pilots undergo the same modulation and SCMA encoding process as data, yielding $\mathbf{x}_{\text{p},k,t}$.}
	\end{remark}
	
	After collecting received measurements $\mathbf{y}_{\text{U},t,n_\text{s}}$ in Eq.~(\ref{yUt}) along $T_2$ time slots, we have
	\begin{equation}\label{YU}
		\mathbf{Y}_{\text{U},n_\text{s}} = \sum\limits_{k=1}^K \sum\limits_{i=1}^2 \mathbf{H}_{\text{r},i} \text{diag}(\mathbf{h}_{\text{b},ik}) \boldsymbol{\Theta}_i \mathbf{X}_{k,n_\text{s}} + \mathbf{N}_{\text{U},n_\text{s}},
	\end{equation}
	where $\mathbf{X}_{k,n_\text{s}} = \text{diag}(x_{k,1,n_\text{s}},...,x_{k,T_2,n_\text{s}})$, $\mathbf{N}_U = [\mathbf{n}_{\text{U},1,n_\text{s}},...,\\\mathbf{n}_{\text{U},T_2,n_\text{s}}]$. 
	
	In the following sections, we present an SCMA-based data detection and UAMP-SBL-based channel sensing method, which can yield initial estimates of channel angles and data symbols. Then, we propose an iterative framework that utilizes reduced-dimension matrix calculations to refine these estimates.
	
	\subsection{SCMA-UAMP-SBL for initial estimates}
	We first recover the coarse estimate $\mathbf{\hat{h}}_{\text{b},ik}^{(0)}$ from sensing pilots by the LSE. The detailed update process can be found in the Appendix~\ref{h^0}. Then, the estimation for cascaded channels of the $k$-th UE can be written as $\mathbf{\hat{H}}_{\text{casc},k}^{(0)} = \sum_{i=1}^2 \mathbf{H}_{\text{r},i} \text{diag}(\mathbf{\hat{h}}_{\text{b},ik}^{(0)}) \boldsymbol{\Theta}_i$. Thus, the received signal minus the estimated pilot part can be written as $\mathbf{\hat{Y}}_{\text{d},n_\text{s}}^{(0)} = \mathbf{Y}_{\text{U},n_\text{s}} - \sum_{k=1}^K \mathbf{\hat{H}}_{\text{casc},k}^{(0)} \mathbf{X}_{\text{p},k,n_\text{s}}$, which can be viewed as the estimate of the received signal from data symbols, and further be written as
	\begin{equation}\label{Ydns}
		\mathbf{\hat{Y}}_{\text{d},n_\text{s}}^{(0)} = \sum\limits_{k \in \mathbf{F}_{n_\text{s}}} \mathbf{\hat{H}}_{\text{casc},k}^{(0)} \mathbf{X}_{\text{d},k,n_\text{s}}+\mathbf{N}_{\text{equ},n_\text{s}},
	\end{equation}
	where $\mathbf{F}_{n_\text{s}}$ is the set of collision UEs in $n_\text{s}$-th subcarrier and $\mathbf{N}_{\text{equ},n_\text{s}}$ is assumed the equivalent error of the received signal having the Gaussian distribution with the variance of $N_0$.

	To reduce the complexity of maximum likelihood (ML) decoders, we obtain data estimates in an iterative way by message passing (MP) methods. Furthermore, in order to efficiently estimate transmission codewords, we modify traditional SCMA decoders to further utilize the diversity gain from multiple-antenna resources. For this purpose, the time resource for data detection can be expanded to $MT_2$ dimension. On the basis of the above two points, (\ref{Ydns}) can be rewritten as
	\begin{equation}
		y_{t,n_\text{s}} = \sum\limits_{k \in \mathbf{F}_{n_\text{s}}} h_{k,t} x_{\text{d},k,t,n_\text{s}} + n_{t,n_\text{s}},
	\end{equation}
	where $y_{t,n_\text{s}}$ and $h_{k,t}$ are the $m$-th row and $t_2$-th column element of $\mathbf{\hat{Y}}_{\text{d},n_\text{s}}^{(0)} \in \mathbb{C}^{M \times T_2}$ and $\mathbf{\hat{H}}_{\text{casc},k}^{(0)} \in \mathbb{C}^{M \times T_2}$, resepctively, with $m=\text{mod}(t,M)$ and $t_2=\lceil t/M \rceil$.   
	
	\subsubsection{Proposed SCMA decoder for data detection}
	The decoding process is to calculate the probability of the $n_{\text{c}}$-th codeword for the $k$-th UE, and then pick the codeword with the maximum posterior probability among $n_{\text{c}}\in\{1,...,N_{\text{c}}\}$. This process can perform independent and parallel decoding for different $t$. Let $f_{n_\text{s}}$ and $x_{k,n_\text{s}}$ represent the function node and variable node in the factor graph for $y_{t,n_\text{s}}$ and $x_{\text{d},k,t,n_\text{s}}$, respectively, where the subscript $t$ is omitted. Based on the MP updating rule, the message sent from function node $f_{n_\text{s}}$, to variable node $x_{k,n_\text{s}}$ can be written as
	\begin{equation}\label{f2v}
		\begin{aligned}
			&I_{f_{n_\text{s}} \to x_{k,n_\text{s}}}^{(s)}(\mathbf{C}_{k,n_\text{s}}(n_{\text{c}})) =\!\!\!\! \sum\limits_{\boldsymbol{\varepsilon}_{l_k,n_\text{s}},l_k \in \mathbf{F}_{n_\text{s}} \setminus k}\!\!\!\! f_{n_\text{s}}(\boldsymbol{\varepsilon}_{l_k,n_\text{s}},\mathbf{C}_{k,n_\text{s}}(n_{\text{c}}))\\
			&  \prod\limits_{l_k\in \mathbf{F}_{n_\text{s}} \setminus k} I_{x_{l_k,n_\text{s}} \to f_{n_\text{s}} }^{(s-1)}(\mathbf{C}_{l_k,n_\text{s}}(n_{c})), \quad \text{for} \ \forall n_{\text{c}},  \ \forall k \in \mathbf{F}_{n_\text{s}},   
		\end{aligned}
	\end{equation}
	where the likelihood function $f_{n_\text{s}}(\mathbf{x}_{n_\text{s}})$ is given by
	\begin{equation}
		f_{n_\text{s}}(\mathbf{x}_{\text{d},k,t}) \propto \text{exp}\big\{ -\frac{1}{N_0} |{y}_{t,n_\text{s}}- \sum\limits_{k \in \mathbf{F}_{n_\text{s}}} h_{k,t} {x}_{\text{d},k,t,n_\text{s}}|^2  \big\},
	\end{equation} 
	$\mathbf{C}_{k,n_\text{s}}(n_{\text{c}})$ is the $n_{\text{c}}$-th codeword for the $k$-th UE and $n_\text{s}$-th frequency band, and $\boldsymbol{\varepsilon}_{l_k,n_\text{s}}=[\mathbf{C}_{l_k,n_\text{s}}(n_{c})]_{l_k \in \mathbf{F}_{n_\text{s}} \setminus k}$, $I_{{x}_{l_k,n_\text{s}} \to f_{n_\text{s}} }^{(s-1)}(\mathbf{C}_{l_k,n_\text{s}}(n_{c}))$ is the extrinsic message passed from nodes $\mathbf{C}_{l_k,n_\text{s}}(n_{c})$ to $f_{n_\text{s}}$ at the $(s-1)$-th iteration and can be updated by
	\begin{equation}\label{v2f}
		\begin{aligned}
			&I_{{x}_{k,n_\text{s}} \to f_{n_\text{s}} }^{(s)}(\mathbf{C}_{k,n_\text{s}}(n_{\text{c}})) = p_{{x}_{k,n_\text{s}}}^{{(s)}} \cdot \\
			& \prod\limits_{m_s \in \mathbf{G}_{k} \setminus n_\text{s}} I_{f_{m_s} \to {x}_{k,m_s}}^{(s)}(\mathbf{C}_{k,m_s}(n_{\text{c}})),\  \text{for} \ \forall n_{\text{c}}, \ \forall k, \ \forall n_\text{s} \in \mathbf{G}_k,
		\end{aligned}
	\end{equation}
	where $\mathbf{G}_k$ represents subcarrier indexes corresponding to the $k$-th UE, and $p_{{x}_{k,n_\text{s}}}^{(s)} = \big(\sum\limits_{n_{\text{c}}=1}^{N_{\text{c}}}  I_{{x}_{k,n_\text{s}} \to f_{n_\text{s}} }^{(s-1)}(\mathbf{C}_{k,n_\text{s}}(n_{\text{c}})) \big)^{-1}$.
	
	This iterative process can be initialized by assuming equal prior probabilities for each codeword, i.e., $I_{{x}_{k,n_\text{s}} \to f_{n_\text{s}}}^{(0)}(\mathbf{C}_{k,n_\text{s}}(n_{c})) = \frac{1}{N_{\text{c}}}, \ \text{for} \  \forall k, \ \forall n_\text{s},\ \forall n_{\text{c}}$.

	Suppose that the number of iterations at convergence is $S$, and the estimated data symbols $\mathbf{X}_{\text{d},k,n_\text{s}}^{(1)}$ can be determined by the chosen transmission codewords expressed as
	\begin{equation}\label{Ck}
		\mathbf{\hat{C}}_k = \arg \max\limits_{n_{\text{c}}=1,...,N_{\text{c}}} \prod\limits_{m_s \in G_k} I_{f_{m_s} \to {x}_{k,m_s}}^{(S)}(\mathbf{C}_{k,m_s}(n_{\text{c}})).
	\end{equation}
	
	\begin{remark}
		\emph{Each iteration of the MP method involves calculating the product of a large number of small probability values, which often causes numerical precision underflow and ultimately impairs the convergence results. This challenge can be effectively resolved by using log-likelihood probabilities, which convert the original multiplication operations to addition operations. Then, we calculate the log-likelihood rate to determine the estimated codewords.}
	\end{remark}
	
	\subsubsection{Proposed UAMP-SBL method for sparse vector recovery}
	Based on data estimates in (\ref{Ck}), we further estimate the sparse vector $\boldsymbol{\psi}$. Considering the vectorization $\mathbf{y}_{\text{U},n_\text{s}} = \text{vec}(\mathbf{Y}_{\text{U},n_\text{s}})$ in Eq.~(\ref{YU}), and employing the property $\text{vec}(\mathbf{A} \text{diag}(\mathbf{b}) \mathbf{C}) = (\mathbf{C}^T \odot \mathbf{A}) \mathbf{b}$, where $\odot$ represents the Khatri-Rao product, we have
	\begin{equation}
		\mathbf{y}_{\text{U},n_\text{s}} = \sum\limits_{k=1}^K \sum\limits_{i=1}^2 \left( (\boldsymbol{\Theta}_i \mathbf{X}_{k,n_\text{s}}^{(1)} )^T \odot \mathbf{H}_{\text{r},i} \right) \mathbf{h}_{\text{b},ik} + \mathbf{n}_{\text{U},n_\text{s}},
	\end{equation}
	
	By introducing the mmWave sparse model of $\mathbf{h}_{\text{b},ik}$ in Eq.~(\ref{hb_sparse}), we have
	\begin{equation}\label{yU}
		\mathbf{y}_{\text{U},n_\text{s}} = \mathbf{V}(\mathbf{X}_{n_\text{s}}^{(1)}) \mathbf{B}_{\text{U}2}(\mathbf{w}_{\text{\text{U2R}},i}^{\text{A}}, \mathbf{g}_{\text{\text{U2R}},i}^{\text{A}}) \boldsymbol{\psi}+ \mathbf{n}_{\text{U},n_\text{s}},
	\end{equation}
	where $\mathbf{V}(\mathbf{X}_{n_\text{s}}^{(1)}) = \big[\mathbf{V}_1(\mathbf{X}_{1,n_\text{s}}^{(1)}),...,\mathbf{V}_K(\mathbf{X}_{K,n_\text{s}}^{(1)})  \big]\in  \mathbb{C}^{MT_2\times 2KN}$, $\mathbf{V}_k(\mathbf{X}_{k,n_\text{s}}^{(1)}) = [\sqrt{N_1} \left( (\boldsymbol{\Theta}_1 \mathbf{X}_{k,n_\text{s}}^{(1)} )^T \odot \mathbf{H}_{\text{r},1} \right) \\, \sqrt{N_2} ( (\boldsymbol{\Theta}_2 \mathbf{X}_{k,n_\text{s}}^{(1)} )^T \odot \mathbf{H}_{\text{r},2} ) ]$, $\mathbf{B}_{\text{U}2}(\mathbf{w}_{\text{\text{U2R}},i}^{\text{A}}, \mathbf{g}_{\text{\text{U2R}},i}^{\text{A}}) = \text{blkdiag}\big{\{}  \mathbf{B}_{\text{U}}(\mathbf{w}_{\text{\text{U2R}},i}^{\text{A}}, \mathbf{g}_{\text{\text{U2R}},i}^{\text{A}}),..., \mathbf{B}_{\text{U}}(\mathbf{w}_{\text{\text{U2R}},i}^{\text{A}}, \mathbf{g}_{\text{\text{U2R}},i}^{\text{A}})\big{\}} \in \mathbb{C}^{2KN \times 2K\text{G}_4\text{G}_5}$, $\mathbf{B}_{\text{U}}(\mathbf{w}_{\text{\text{U2R}},i}^{\text{A}}, \mathbf{g}_{\text{\text{U2R}},i}^{\text{A}}) = \text{blkdiag}\big{\{}  \mathbf{B}_{\text{\text{U2R}},1}(\mathbf{w}_{\text{\text{U2R}},1}^{\text{A}}, \mathbf{g}_{\text{\text{U2R}},1}^{\text{A}}), \ \mathbf{B}_{\text{\text{U2R}},2}(\mathbf{w}_{\text{\text{U2R}},2}^{\text{A}}, \mathbf{g}_{\text{\text{U2R}},2}^{\text{A}})\big{\}} $, blkdiag$(\cdot)$ is the block diagonalization operation, $\boldsymbol{\psi} = [\boldsymbol{\psi}_1^T,...,\boldsymbol{\psi}_K^T]^T$ and $\boldsymbol{\psi}_k = [\boldsymbol{\psi}_{1,k}^T, \ \boldsymbol{\psi}_{2,k}^T]^T$.
	
	Then, we collect (\ref{yU}) in $N_\text{s}$ frequency bands to form
	\begin{equation}\label{yU}
		\mathbf{y}_U = \mathbf{V}_2(\mathbf{X}^{(1)}) \mathbf{B}_{\text{U}2}(\mathbf{w}_{\text{\text{U2R}},i}^{\text{A}}, \mathbf{g}_{\text{\text{U2R}},i}^{\text{A}}) \boldsymbol{\psi}+ \mathbf{n}_{\text{U}},
	\end{equation}
	where $\mathbf{y}_{\text{U}} = [\mathbf{y}_{\text{U},1}^T,...,\mathbf{y}_{\text{U},N_\text{s}}^T]^T \in \mathbb{C}^{MT_2N_\text{s} \times 1}$, $\mathbf{V}_2(\mathbf{X}^{(1)}) = [\mathbf{V}^T(\mathbf{X}_1^{(1)}),...,\mathbf{V}^T(\mathbf{X}_{N_\text{s}}^{(1)})]^T\in \mathbb{C}^{MT_2N_\text{s} \times 2KN}$ and $\mathbf{n}_U = [\mathbf{n}_{\text{U},1}^T,...,\mathbf{n}_{\text{U},N_\text{s}}^T]^T$. In this way, the received signals at the BS from multiple UEs at all time and frequency resourse blocks are composed into a standard compressed sensing vector form, where $\mathbf{D}(\mathbf{X}^{(1)}, \mathbf{w}_{\text{\text{U2R}},i}^{\text{A}}, \mathbf{g}_{\text{\text{U2R}},i}^{\text{A}}) =   \mathbf{V}_2(\mathbf{X}^{(1)}) \mathbf{B}_{\text{U}2}(\mathbf{w}_{\text{\text{U2R}},i}^{\text{A}}, \mathbf{g}_{\text{\text{U2R}},i}^{\text{A}}) \in \mathbb{C}^{MT_2N_\text{s} \times 2K\text{G}_4\text{G}_5}$ is the sensing matrix and $\boldsymbol{\psi}$ is the sparse vector. Specifically, the sensing matrix is a function w.r.t superimposed symbols $\mathbf{X}^{(1)}$ and selected angle grids $\{\mathbf{w}_{\text{\text{U2R}},i}^{\text{A}}, \mathbf{g}_{\text{\text{U2R}},i}^{\text{A}}\}_{i=1}^2$. Thus, this ISAC problem can be solved by using a similar framework as in Section \ref{sec_fix}. However, the high complexity of SBL is a concern in practical ISAC applications. To address this issue, we employ UAMP-SBL to avoid the high-complexity inversion operations.
	
	It is shown in \cite{Guo2015} that the robustness of AMP is remarkably improved through simple pre-processing, i.e., performing a unitary transformation to the original formulation. By performing singular value decomposition on $\mathbf{D}(\mathbf{X}^{(1)}, \mathbf{w}_{\text{\text{U2R}},i}^{\text{A}}, \mathbf{g}_{\text{\text{U2R}},i}^{\text{A}})$ with $\mathbf{U}_L \boldsymbol{\Sigma}_D \mathbf{U}_R^H$, we multiply both sides of Eq.~(\ref{yU}) by $\mathbf{U}_L^H$ as
	\begin{equation}
		\mathbf{r} = \boldsymbol{\Lambda} \boldsymbol{\psi} + \bar{\boldsymbol{n}}_U,
	\end{equation}
	where $\mathbf{r} = \mathbf{U}_L^H \mathbf{y}_U$, $\boldsymbol{\Lambda} = \boldsymbol{\Sigma}_D \mathbf{U}_R^H$ and $\bar{\boldsymbol{n}}_U = \mathbf{U}_L^H \mathbf{n}_U$ still follows $\mathcal{CN}(\mathbf{0},N_0 \mathbf{I})$. Similar with the framework of SBL, we assume for $g \in \{1,...,2K\text{G}_4\text{G}_5\}$ that
	\begin{equation}
		p(\psi_{g}| \gamma_{{\psi},g}) = \mathcal{CN}(\psi_g;0,\gamma_{{\psi},g}^{-1}).
	\end{equation}
	
	The adaptive $\gamma_g$ follows Gamma distribution, i.e., $p(\gamma_{{\psi},g} | a_{\psi}, b_{\psi}) = \Gamma(\gamma_{{\psi},g};1+a_{\psi}, b_{\psi})$. Then, we have $p(r_q|\boldsymbol{\psi}, \beta_{\psi}) = \mathcal{CN}(r_q;  \boldsymbol{\Lambda}(q,:) \boldsymbol{\psi}, \beta_{\psi}^{-1})$, where $\beta_{\psi}$ is the noise precision parameter, $\boldsymbol{\Lambda} (q,:)$ is the $q$-th row of matrix $\boldsymbol{\Lambda}$. Then introduce hard constraint factors $p(\kappa_q | \boldsymbol{\psi}) = \delta(\kappa_q- \boldsymbol{\Lambda} (q,:) \boldsymbol{\psi})$ with $\boldsymbol{\kappa} = \boldsymbol{\Lambda} \boldsymbol{\psi}$ to facilitate the incorporation of UAMP in the SBL, which is crucial to an efficient realization of SBL. Then, the conditional joint distribution is
	\begin{equation}\label{facto}
		\begin{aligned}
			&p(\boldsymbol{\psi}, \boldsymbol{\kappa}, \boldsymbol{\gamma}_{\psi}, \beta_{\psi} | \mathbf{r})  \propto p(\mathbf{r}|\boldsymbol{\kappa}, \beta_{\psi}) p(\boldsymbol{\kappa}|\boldsymbol{\psi}) p(\boldsymbol{\psi}| \boldsymbol{\gamma}_{\psi}) p(\boldsymbol{\gamma}_{\psi} | a_{\psi}, b_{\psi}) \\
			&\quad \cdot p(\beta_{\psi}) = \prod\limits_{q=1}^{MT_2N_\text{s}} \mathcal{CN}(r_q| \kappa_q, \beta_{\psi}^{-1}) \prod\limits_{q=1}^{MT_2N_\text{s}} \delta(\kappa_q - \boldsymbol{\Lambda} (q,:) \boldsymbol{\psi}) \\
			&\cdot \prod\limits_{g=1}^{2K\text{G}_4\text{G}_5}\mathbf{CN}(\psi_g|0,\gamma_{{\psi},g}^{-1}) \prod\limits_{g=1}^{2K\text{G}_4\text{G}_5}\Gamma(\gamma_{{\psi},g}| 1+a_{\psi}, b_{\psi}) p({\beta_{\psi}}).
		\end{aligned}
	\end{equation}
	
\begin{figure}[ht]
		\centering
		\includegraphics[scale=0.3]{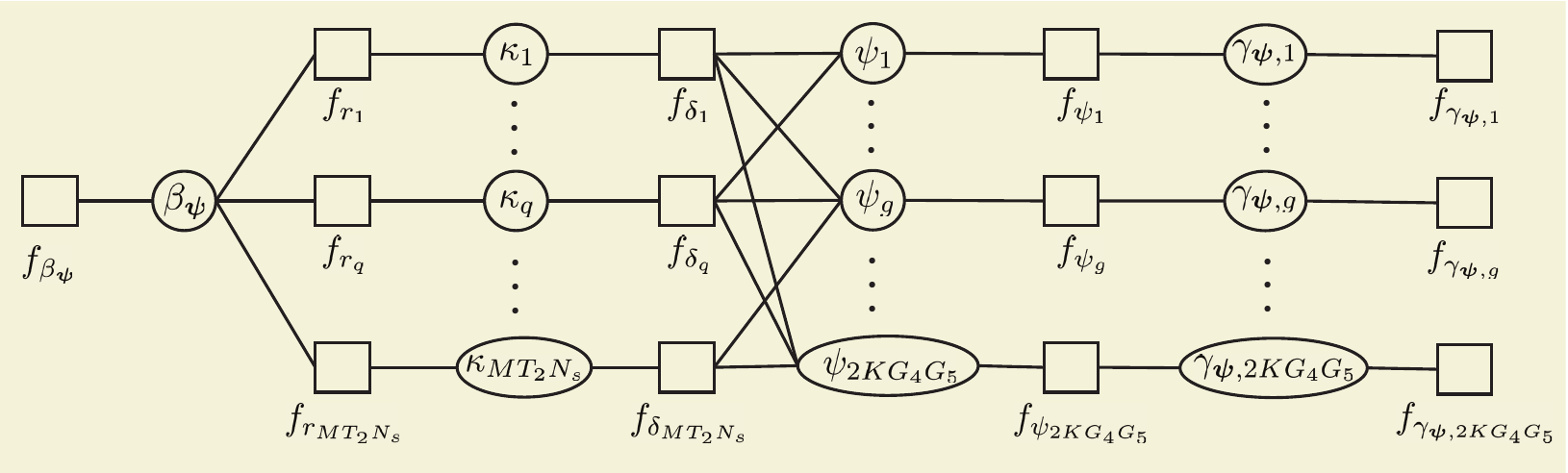}
		\caption{Factor graph of (\ref{facto}) for deriving UAMP-SBL.}
		\label{utamp}
	\end{figure}

	To facilitate the derivation of the MP algorithm, a factor graph representation of the factorization in (\ref{facto}) is shown in Fig.~\ref{utamp}. The local functions are denoted  by $f_{\beta_{\psi}}(\beta_{\psi}) \propto 1/\beta_{\psi}$, $f_{\text{r}_{q}}(r_{q}, \kappa_{q}, \beta_{\psi}) = \mathcal{CN}(r_{q}| \kappa_{q}, \beta_{\psi}^{-1})$, $f_{\delta_{q}} = \delta(\kappa_{q}- \boldsymbol{\Lambda}(q,:) \boldsymbol{\psi})$, $f_{\psi_g}(\psi_g, \gamma_{\psi}) = \mathcal{CN}(\psi_g | 0,\beta_{\psi}^{-1})$ and $f_{\gamma_{{\psi},g}} = \Gamma(\gamma_{{\psi},g}| 1+a_{\psi}, b_{\psi}) p({\beta_{\psi}})$. Based on the work in \cite{uamp} and the graph representation, the UAMP-SBL algorithm can be adapted to obtain $\boldsymbol{\psi}^{(1)}$ efficiently, which is summarized in Algorithm~\ref{stage2_DoA}.
	
	\begin{algorithm}
		\caption{UAMP-SBL \cite{uamp} based method to obtain $\boldsymbol{\psi}$}
		\label{stage2_DoA}
		\begin{algorithmic}[1]
			\STATE Unitary transform: $\mathbf{r} = \mathbf{U}_L^H \mathbf{y}_U$ and $\boldsymbol{\Lambda} = \boldsymbol{\Sigma}_D \mathbf{U}_R^H$.
			\STATE Initialize $\boldsymbol{\lambda}_e = \boldsymbol{\Sigma}_D \boldsymbol{\Sigma}_D^H \mathbf{1}$, $\tau_{\psi}^{(0)}=1$, $\boldsymbol{\hat{\psi}}^{(0)}=\mathbf{0}$, $a_\epsilon=b_{\epsilon}=10^{-3}$, $\boldsymbol{\hat{\gamma}}_{\psi}=\mathbf{1}$, $\hat{\beta}_{\psi}=1$, $\mathbf{s}=\mathbf{0}$ and the iteration number $u=0$.
			\REPEAT
			\STATE $\boldsymbol{\tau}_p = \tau_{\psi}^{(u)} \boldsymbol{\lambda}_e$
			\STATE $\mathbf{p}=\boldsymbol{\Lambda} \hat{\boldsymbol{\psi}}^{(u)} - \boldsymbol{\tau}_p \mathbf{s}$
			\STATE $\mathbf{v}_{\boldsymbol{\kappa}} = \boldsymbol{\tau}_p ./ (1+\hat{\beta}_{\psi} \boldsymbol{\tau}_p)$
			\STATE $\hat{\boldsymbol{\kappa}} = (\hat{\beta}_{\psi} \boldsymbol{\tau}_p \cdot \mathbf{r}+ \mathbf{p}) ./ (\mathbf{1}+ \hat{\beta}_{\psi} \boldsymbol{\tau}_p)$
			\STATE $\hat{\beta}_{\psi} = MT_2N_\text{s}/(\|\mathbf{r}-\boldsymbol{\kappa}\|^2+ \mathbf{1}^H \mathbf{v}_\kappa)$
			\STATE $\boldsymbol{\tau}_s=\mathbf{1}./(\boldsymbol{\tau}_p + \hat{\beta}_{\psi}^{-1} \mathbf{1})$
			\STATE $\mathbf{s} = \boldsymbol{\tau}_s \cdot (\mathbf{r}-\mathbf{p})$
			\STATE $\tau_q^{-1} = (1/(2K\text{G}_4\text{G}_5)) \boldsymbol{\lambda})_e ^H \boldsymbol{\tau}_s$
			\STATE $\mathbf{q}=\hat{\boldsymbol{\psi}}^{(u)} + \tau_q \boldsymbol{\Lambda}^H \mathbf{s}$
			\STATE $\tau_\psi^{(j+1)} = (\tau_q /(2K\text{G}_4\text{G}_5)) \mathbf{1}^H (\mathbf{1}./(\mathbf{1}+ \tau_q \hat{\boldsymbol{\gamma}}_\psi))$
			\STATE $\hat{\boldsymbol{\psi}}^{(u+1)} = \mathbf{q}./(\mathbf{1}+ \tau_q \hat{\boldsymbol{\gamma}}_\psi)$
			\STATE $\hat{\boldsymbol{\gamma}}_\psi(g) = (2a_\epsilon +1)/(b_{\psi}+|\hat{\boldsymbol{\psi}}^{(u+1)}(g) |^2+ \tau_\psi^{(j+1)})$, $\forall g$
			\STATE $a_\epsilon = \scriptstyle \frac{1}{2} \sqrt{\log (\frac{1}{2K\text{G}_4\text{G}_5} \sum_{g=1}^{2K\text{G}_4\text{G}_5} \hat{\boldsymbol{\gamma}}_\psi(g) ) - \frac{1}{N} \sum_{g=1}^{2K\text{G}_4\text{G}_5} \log(\hat{\boldsymbol{\gamma}}_\psi(g) ) }$
			\STATE $u \gets u+1	$
			
			\UNTIL $\big( \| \hat{\boldsymbol{\psi}}^{(u+1)} - \hat{\boldsymbol{\psi}}^{(u)} \|^2 / \|\hat{\boldsymbol{\psi}}^{(u+1)} \|^2 < \delta_\psi $ or $u>u_{\max} \big)$ 
			
		\end{algorithmic}  
	\end{algorithm}
	
	\subsubsection{Sub-gradient descent method for angle grids refinements}
	Based on obtained $\mathbf{X}^{(1)}$ and $\boldsymbol{\psi}^{(1)}$, we treat $\boldsymbol{\eta}=[\mathbf{w}_{\text{\text{U2R}},i}^{\text{A}}, \mathbf{g}_{\text{\text{U2R}},i}^{\text{A}}]_{i=1}^2$ as unknown parameters and formulate the following off-grid problem
	\begin{equation}\label{eta}
		\arg \min_{\boldsymbol{\eta}} \|\mathbf{y}_U -  \mathbf{V}_2 \mathbf{B}_{\text{U}2}(\boldsymbol{\eta}) \boldsymbol{\psi}^{(1)} \|^2,
	\end{equation} 
	which is the LSE of (35) and $\mathbf{X}^{(1)}$ is omitted here. Similarly, it is difficult to find the closed-form solution with $\boldsymbol{\eta}$ to solve the problem (\ref{eta}) because of the nonlinear relationship of $\mathbf{B}_{\text{U}2}(\boldsymbol{\eta})$ function w.r.t. $\boldsymbol{\eta}$. Hence, we resort to the sub-gradient descent method \cite{stepsize} with adjustable stepsize as $\boldsymbol{\eta}^{(j+1)} = \boldsymbol{\eta}^{(j)} + \epsilon_2^{(j)}  \mathcal{F}_2(\boldsymbol{\eta}^{(j)})$ and $\epsilon_2^{(j)} = \frac{\eta_{\max} - \eta_{\min}}{c_2 (1+d_2 \cdot j)}$, where $\mathcal{F}_2(\boldsymbol{\eta}) = 2 \text{Re}\left( \boldsymbol{\psi}^H \mathbf{B}_{\text{U}2}^H(\boldsymbol{\eta}) \mathbf{V}_2^H \mathbf{V}_2 \frac{\partial \mathbf{B}_{\text{U}2}(\boldsymbol{\eta})}{\partial \boldsymbol{\eta}} \boldsymbol{\psi} - \mathbf{y}_U^H \mathbf{V}_2 \frac{\partial \mathbf{B}_{\text{U}2}(\boldsymbol{\eta})}{\partial \boldsymbol{\eta}} \boldsymbol{\psi} \right)$ and can be obtained in closed forms, $\eta_{\max}$ and $\eta_{\min}$ are the maximum and minmum values of the selected grids, respectively, $c_2$ and $d_2$ are learning rate and decay coefficient.

	\subsection{Estimate refinements by reduced-dimension matrix calculations after the first iteration}
	To further reduce the complexity of the proposed multi-UE simultaneous communication and localization algorithm in subsequent iterations, we can reduce the dimensions of the matrix calculations involved in updating $\boldsymbol{\psi}$ and $\boldsymbol{\eta}$. This can be performed for the $j$-th iteration by rewriting (\ref{yU}) as follows
	\begin{equation}\label{yu_re}
		\mathbf{{y}}_U = \mathbf{D}_{\text{redu}}(\mathbf{X}^{(j)}, \boldsymbol{\breve{\eta}}^{(j)} ) \boldsymbol{\breve{\psi}}^{(j)} + \mathbf{n}_U.
	\end{equation}
	
	Such a conversion is due to the fact that most elements in $\boldsymbol{\psi}$ are close to zero. Therefore, we can extract the elements with large values in $\boldsymbol{\psi}$ to recompose a reduced-dimension vector $\boldsymbol{\breve{\psi}}$ for the $j$-th iteration ($j>1$). Accordingly, the selected rows of $\boldsymbol{\psi}$ further determine the corresponding columns of $\mathbf{D}(\mathbf{X}^{(j)}, \boldsymbol{\eta}^{(j)})$, which then yields $\mathbf{D}_{\text{redu}}(\mathbf{X}^{(j)}, \boldsymbol{\breve{\eta}}^{(j)} )$, and $\boldsymbol{\breve{\eta}}^{(j)}$ can be viewed as the selected angle grids.
	\begin{remark}
		\emph{Note that $\boldsymbol{\psi} \in \mathbb{C}^{2K\text{G}_4\text{G}_5 \times 1}$ is composed of $\boldsymbol{\psi}_{i,k} \in \mathbb{C}^{\text{G}_4\text{G}_5 \times 1}$ for the $i$-th RIS and $k$-th UE. Therefore, the selection criteria for $\boldsymbol{\breve{\psi}}$ should be considered separately for each $\boldsymbol{\psi}_{i,k}$ range, ensuring that the finest grid $\boldsymbol{\breve{\psi}}_{i,k}$ corresponding to the effective AoA at the $i$-th RIS from the $k$-th UE is selected. This reduces the $2K\text{G}_4\text{G}_5$-dimension vector to a $2K$-dimension one, significantly lowering the complexity, especially for large $\text{G}_4$ and $\text{G}_5$. Furthermore, the model mismatch error caused by the dimension reduction can be effectively solved using a similar approach to Remark 2.} 
	\end{remark}
	
	Details for updating $\mathbf{X}_d^{(j+1)}$, $\boldsymbol{\breve{\psi}}^{(j+1)}$ and $\boldsymbol{\breve{\eta}}^{(j+1)}$ are presented as follows:
	
	\textbf{STEP 1} Update $\mathbf{X}_d^{(j+1)}$: The same with the proposed SCMA decoder as in (32).
	
	\textbf{STEP 2} Update $\boldsymbol{\breve{\psi}}^{(j+1)}$: Solving the reduced-dimension problem $\boldsymbol{\breve{\psi}}^{(j+1)} = \arg \min_{\boldsymbol{\breve{\psi}}} \|\mathbf{y}_U-\mathbf{D}_{\text{redu}}(\mathbf{X}^{(j+1)}, \boldsymbol{\breve{\eta}}^{(j)}) \boldsymbol{\breve{\psi}} \|^2 $ by the low-complexity MP method.
	
	\textbf{STEP 3} Update $\boldsymbol{\breve{\eta}}^{(j+1)}$: Similar to the sub-gradient descent method as in (41) by reduced-dimension matrix calculations.
	
	\subsection{Mapping $\boldsymbol{\breve{\eta}}$ to multi-UE location coordinates}
	Suppose that the number of iterations at convergence is $J$, and the final grid estimates $\boldsymbol{\breve{\eta}}^{(J)}$ will map to estimated locations of multi-UE in the following derivation. Denote the coordinate of the $i$-th RIS and $k$-th UE by $\mathbf{q}_{\text{R}i}=[q_{\text{R}i,x}, q_{\text{R}i,\text{y}}, q_{\text{R}i,\text{z}}]$ and $\mathbf{p}_{\text{U}k} = [p_{\text{U}k,x}, p_{\text{U}k,\text{y}}, p_{\text{U}k,\text{z}}]$, respectively, for $\forall i, k$. Owing to the fixed positions of RISs, here $\mathbf{q}_{\text{R}1}$ and $\mathbf{q}_{\text{R}2}$ are all prior known. Then, the estimates of effective AoAs at the RIS-$i$ from the UE-$k$ can be obtained by $\mathbf{\breve{w}}_{\text{U2R},i}(k)$ and $\mathbf{\breve{g}}_{\text{U2R},i}(k)$ for $\forall k$. Let $d_{\text{\text{U2R}},ik}$ denote the distance from the UE-$k$ to the RIS-$i$ with $d_{\text{\text{U2R}},ik} = \|\mathbf{p}_{\text{U}k}-\mathbf{q}_{\text{R}i}\|$. Then, the effective AoAs at the two RIS from UE-$k$ are given below
	\begin{equation}
		\hat{v}_{\text{\text{U2R}},1k}^{\text{A}}= \frac{q_{\text{R}1,\text{z}}-p_{\text{U}k,\text{z}}}{d_{\text{\text{U2R}},1k}}, \quad \hat{u}_{\text{\text{U2R}},1k}^{\text{A}}= \frac{q_{\text{R}1,\text{y}}-p_{\text{U}k,\text{y}}}{d_{\text{\text{U2R}},1k}}
	\end{equation}
	\begin{equation}
		\hat{v}_{\text{\text{U2R}},2k}^{\text{A}}= \frac{q_{\text{R}2,\text{z}}-p_{\text{U}k,\text{z}}}{d_{\text{\text{U2R}},2k}}, \quad \hat{u}_{\text{\text{U2R}},2k}^{\text{A}} = \frac{q_{\text{R}2,\text{y}}-p_{\text{U}k,\text{y}}}{d_{\text{\text{U2R}},2k}}
	\end{equation}
	
	The solution can be expressed as
	\begin{equation}
		p_{\text{U}k,\text{y}} = q_{\text{R}1,\text{y}}-\hat{u}_{\text{\text{U2R}},1k}^{\text{A}} d_{\text{\text{U2R}},1k}
	\end{equation} 
	\begin{equation}
		p_{\text{U}k,\text{z}} = q_{\text{R}1,\text{z}}-\hat{v}_{\text{\text{U2R}},1k}^{\text{A}} d_{\text{\text{U2R}},1k}
	\end{equation}
	\begin{equation}
		d_{\text{\text{U2R}},1k} = \frac{\hat{u}_{\text{\text{U2R}},2k}^{\text{A}} (q_{\text{R}1,\text{z}}-q_{\text{R}2,\text{z}})-\hat{v}_{\text{\text{U2R}},2k}^{\text{A}}(q_{\text{R}1,\text{y}}-q_{\text{R}2,\text{y}})}{\hat{u}_{\text{\text{U2R}},2k}^{\text{A}} \hat{v}_{\text{\text{U2R}},1k}^{\text{A}} -\hat{u}_{\text{\text{U2R}},1k}^{\text{A}} \hat{v}_{\text{\text{U2R}},2k}^{\text{A}} }
	\end{equation}
	\begin{equation}
		d_{\text{\text{U2R}},2k} = \frac{\hat{u}_{\text{\text{U2R}},1k}^{\text{A}} (q_{\text{R}1,\text{z}}-q_{\text{R}2,\text{z}})-\hat{v}_{\text{\text{U2R}},1k}^{\text{A}} (q_{\text{R}1,\text{y}}-q_{\text{R}2,\text{y}})}{\hat{u}_{\text{\text{U2R}},2k}^{\text{A}} \hat{v}_{\text{\text{U2R}},1k}^{\text{A}} -\hat{u}_{\text{\text{U2R}},1k}^{\text{A}} \hat{v}_{\text{\text{U2R}},2k}^{\text{A}}}
	\end{equation}
	
	Since the BS and UEs lie on the same side of reflective RISs, here $p_{\text{U}k,\text{x}}$ can be determined by $\hat{p}_{\text{U}k,\text{x}i} = q_{\text{R}i,\text{x}} + \sqrt{d_{\text{\text{U2R}},ik}^2 - (q_{\text{R}i,\text{y}}-p_{\text{U}k,\text{y}})^2-(q_{\text{R}i,\text{z}}-p_{\text{U}k,\text{z}})^2}$ and $p_{\text{U}k,\text{x}} = (\hat{p}_{\text{U}k,\text{x}1}+\hat{p}_{\text{U}k,\text{x}2})/2$.
	
	\begin{figure}[ht]
		\centering
		\includegraphics[scale=0.3]{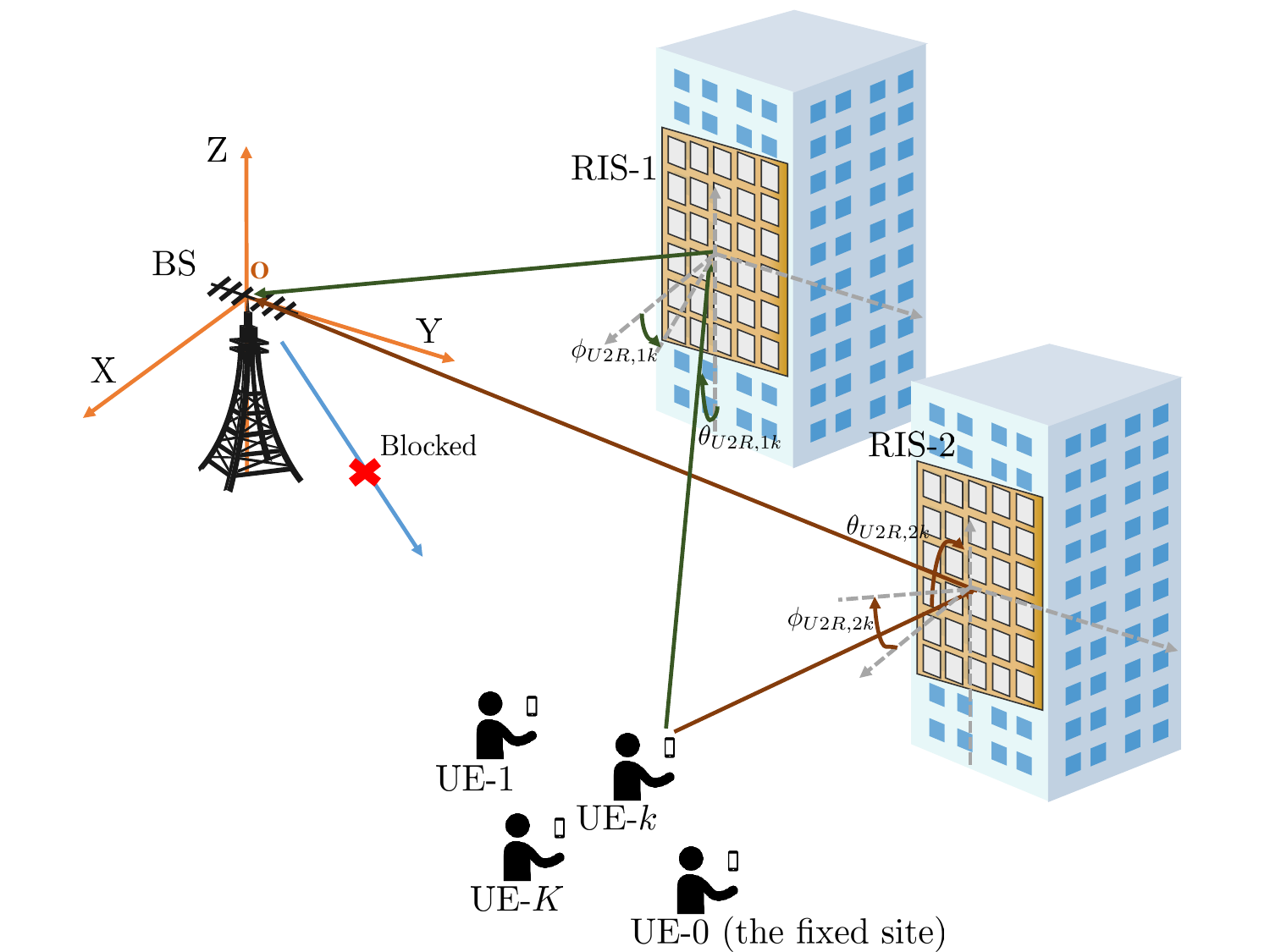}
		\caption{Illustration of geometric relationships of uplink multi-UE ISAC systems.}
		\label{coordinate}
	\end{figure}
	
	\subsection{Summarization and Computational Complexity Analysis}
	\begin{algorithm}
		\caption{The proposed algorithm for multi-UE simultaneous communication and localization}
		\label{localization}
		\begin{algorithmic}[1]
			\STATE Initialize the coarse estimate $\boldsymbol{\psi}^{(0)}$ from sensing pilots.
			\STATE Estimate transmission codewords by the SCMA-MPA decoder and obtain data symbols $\mathbf{X}_{d}^{(1)}$. 
			\STATE Estimate the sparse vector $\boldsymbol{\psi}^{(1)}$ by \emph{Algorithm 1}.
			\STATE Estimate the angle grids $\boldsymbol{\eta}^{(1)}$ by sub-gradient descent method.
			\STATE Initialize the iteration number $j=1$.
			\REPEAT
			\STATE Update $\mathbf{X}_d^{(j+1)}$ based on $\boldsymbol{\psi}^{(j)}$ and $\boldsymbol{\eta}^{(j)}$ by the SCMA-MPA.
			\STATE Update $\boldsymbol{\breve{\psi}}^{(j+1)}$ based on $\mathbf{D}_{\text{redu}}(\mathbf{X}_d^{(j)}, \boldsymbol{\breve{\eta}}^{(j)})$ by the low-complexity MP.
			\STATE Update $\boldsymbol{\breve{\eta}}^{(j+1)}$ based on $\mathbf{X}_d^{(j+1)}$ and $\boldsymbol{\breve{\psi}}^{(j+1)}$ via sub-gradient descent method.
			\STATE $j \gets j+1	$
			
			\UNTIL $\big( \| \boldsymbol{\breve{\eta}}^{(j+1)} - \boldsymbol{\breve{\eta}}^{(j)} \|^2 / \|\boldsymbol{\breve{\eta}}^{(j+1)} \|^2 < \delta_\eta $ or $j>j_{\max} \big)$ 
			\STATE Determine locations for multiple UEs by the estimate $\boldsymbol{\breve{\eta}}$.
			
		\end{algorithmic}  
	\end{algorithm}
	
	We first summarize the proposed algorithm in \emph{Algorithm 3} for multi-UE simultaneous communication and localization. Next, we extend the proposed algorithm to the $N_R$-RIS model ($N_R \ge 2$). In this way, the received signal is written as
	\begin{equation}
		\mathbf{Y}_{\text{U},n_\text{s}} = \sum\limits_{k=1}^K \sum\limits_{i=1}^{N_R} \mathbf{H}_{\text{r},i} \text{diag}(\mathbf{h}_{\text{b},ik}) \boldsymbol{\Theta}_i \mathbf{X}_{k,n_\text{s}} + \mathbf{N}_{\text{U},n_\text{s}},
	\end{equation}
	
	By modifying accordingly the form of Eq.~\eqref{yU}, we have
	\begin{equation}\label{yU2}
		\mathbf{y}_U = \mathbf{V}_2(\mathbf{X}^{(1)}) \mathbf{B}_{\text{U}N_R}(\mathbf{w}_{\text{\text{R2U}},i}^{\text{A}}, \mathbf{g}_{\text{\text{R2U}},i}^{\text{A}}) \boldsymbol{\psi}+ \mathbf{n}_{\text{U}},
	\end{equation}
	where $\mathbf{y}_{\text{U}} = [\mathbf{y}_{\text{U},1}^T,...,\mathbf{y}_{\text{U},N_\text{s}}^T]^T \in \mathbb{C}^{MT_2N_\text{s} \times 1}$, $\mathbf{V}_2(\mathbf{X}^{(1)}) = [\mathbf{V}^T(\mathbf{X}_1^{(1)}),...,\mathbf{V}^T(\mathbf{X}_{N_\text{s}}^{(1)})]^T\in \mathbb{C}^{MT_2N_\text{s} \times N_RKN}$, $\mathbf{B}_{\text{U}N_R}(\mathbf{w}_{\text{\text{U2R}},i}^{\text{A}}, \mathbf{g}_{\text{\text{U2R}},i}^{\text{A}}) = \text{blkdiag}\big{\{}  \mathbf{B}_{\text{U}}(\mathbf{w}_{\text{\text{U2R}},i}^{\text{A}}, \mathbf{g}_{\text{\text{U2R}},i}^{\text{A}}),...,\\ \mathbf{B}_{\text{U}}(\mathbf{w}_{\text{\text{U2R}},i}^{\text{A}}, \mathbf{g}_{\text{\text{U2R}},i}^{\text{A}})\big{\}} \in \mathbb{C}^{N_R KN \times N_R K\text{G}_4\text{G}_5}$, and $\mathbf{n}_U = [\mathbf{n}_{\text{U},1}^T,...,\mathbf{n}_{\text{U},N_\text{s}}^T]^T$. \emph{Algorithm 3} can be utilized to obtain multi-UE communication data $\mathbf{X}_d$ and estimated angle vector $\boldsymbol{\breve{\eta}}$. Then, the estimates of effective AoAs at the RIS-$i$ from the UE-$k$ can be obtained by $\mathbf{\breve{w}}_{\text{U2R},i}(k)$ and $\mathbf{\breve{g}}_{\text{U2R},i}(k)$ for $\forall k$ from $\boldsymbol{\breve{\eta}}$. The detailed steps of how to map effective AoAs to the location coordinates of multiple UEs are provided as follows, where we have
	\begin{equation}\label{multiRIS11}
		\hat{v}_{\text{\text{R2U}},ik}^{\text{A}}= \frac{q_{\text{R}i,\text{z}}-p_{\text{U}k,\text{z}}}{d_{\text{\text{R2U}},ik}}, \quad \hat{u}_{\text{\text{R2U}},ik}^{\text{A}}= \frac{q_{\text{R}i,\text{y}}-p_{\text{U}k,\text{y}}}{d_{\text{\text{R2U}},ik}}
	\end{equation}
	
	The unknown variables in these equations can be written in vector form as $\mathbf{t}_k = [p_{\text{U}k,\text{y}},p_{\text{U}k,\text{z}},d_{\text{\text{R2U}},1k},..., d_{\text{\text{R2U}},N_Rk}]^T  \in \mathbb{R}^{(N_R+2)\times1}$. Transforming the $2N_R$ equations as in Eq.~\eqref{multiRIS11} into the matrix form,
	\begin{small}
		\begin{equation}\label{super1}
			\underbrace{\begin{bmatrix}
					0  & 1 & \hat{v}_{\text{R2U},1k}^{\text{A}} & 0 & \cdots & 0  \\
					1  & 0 & \hat{u}_{\text{R2U},1k}^{\text{A}} & 0 & \cdots & 0 \\
					0  & 1 & 0 & \hat{v}_{\text{R2U},2k}^{\text{A}} & \cdots & 0\\
					1  & 0 & 0 & \hat{u}_{\text{R2U},2k}^{\text{A}} & \cdots & 0\\
					&  & \cdots & \cdots &  &  \\
					0  & 1 & 0 & 0 & \cdots & \hat{v}_{\text{R2U},N_Rk}^{\text{A}} \\
					1  & 0 & 0 & 0 & \cdots & \hat{u}_{\text{R2U},N_Rk}^{\text{A}}
			\end{bmatrix}}_{\mathbf{A}_k}\!\!
			\underbrace{\begin{bmatrix}
					p_{\text{U}k,\text{y}} \\
					p_{\text{U}k,\text{z}}\\
					d_{\text{\text{R2U}},1k} \\
					d_{\text{\text{R2U}},2k} \\
					\cdots \\
					d_{\text{\text{R2U}},N_Rk}
			\end{bmatrix}}_{\mathbf{t}_k}
			\!\! = \!\!
			\underbrace{\begin{bmatrix}
					q_{R1,z} \\
					q_{R1,y} \\
					q_{R2,z} \\
					q_{R2,y} \\
					\cdots \\
					q_{RN_R,z} \\
					q_{RN_R,z}
			\end{bmatrix}}_{\mathbf{b}},
		\end{equation}
	\end{small}
	where $\mathbf{A}_k \in \mathbb{R}^{2N_R \times (N_R+2)}$ is column-full rank. Hence, when $N_R$ is larger than $2$, Eq.~\eqref{super1} belongs to the super-definite equation which can be solved by least squares. Then, the estimation of the user's location can be converted into
	\begin{equation}
		\arg \min_{\mathbf{t}_k} \|\mathbf{A}_k \mathbf{t}_k - \mathbf{b}\|^2.
	\end{equation}
	
	Since the BS and UEs lie on the same side of reflective RISs, here $p_{\text{U}k,\text{x}}$ can be determined by $\hat{p}_{\text{U}k,\text{x}i} = q_{\text{R}i,\text{x}} + \sqrt{d_{\text{\text{R2U}},ik}^2 - (q_{\text{R}i,\text{y}}-p_{\text{U}k,\text{y}})^2-(q_{\text{R}i,\text{z}}-p_{\text{U}k,\text{z}})^2}$ and $p_{\text{U}k,\text{x}} = \sum_{i=1}^{N_R} \hat{p}_{\text{U}k,\text{x}i}/N_R$.

	Finally, the computational complexity is analyzed from the following two parts for the $N_R$-RIS model:
	
	\begin{itemize}
		\item From the perspective of multi-UE localization, we use the UAMP-SBL method for initial estimation and iterative refinements by reduced-dimension matrix calculations. The computational complexity for UAMP-SBL for initialization is $\mathcal{O}((MT_2N_\text{s})^2 N_R K\text{G}_4\text{G}_5)$ and the complexity per iteration is $\mathcal{O}(MT_2N_\text{s}N_R K\text{G}_4\text{G}_5)$. For iteration refinement with the reduced dimension, the complexity per iteration is $\mathcal{O}(MT_2N_\text{s}N_RK)$. Then, we analyze the complexity of using the conventional SBL method and grid refinements for every iteration without reduced-dimension matrix calculations, which produces a total complexity of $\mathcal{O}\bigg(N_R \big(MT_2N_\text{s}(K\text{G}_4\text{G}_5)^2+(MT_2N_\text{s})^2K\text{G}_4\text{G}_5)+(MT_2N_\text{s})^3\big)\bigg)$ per iteration. As $K < \text{G}_4\text{G}_5 \ll MT_2N_\text{s}$, the complexity of our proposed method is significantly lower than that of the conventional off-grid SBL method.
		
		\item From the perspective of multi-UE detection, the computational complexity mainly comes from the proposed MP decoder with $\mathcal{O}(N_\text{s}d_cN_{\text{c}}^{d_c} + KN_{\text{c}})$. Compared to the complexity of the ML decoder with $\mathcal{O}(N_\text{s} N_{\text{c}}^K)$, the complexity of the adopted decoder is much lower as $d_c \ll K$.  
	\end{itemize}

	\section{Numerical Results}\label{simulation}
	This section presents simulation results to demonstrate the effectiveness of the proposed superimposed symbol scheme in double-RIS aided ISAC systems for the fixed site and multi-UE scenarios, respectively. We also evaluate the impact of signal-to-noise ratio (SNR) or $E_b/N_0$, the transmission period length, and the power proportion factor on proposed algorithms, and compare them with other benchmarks.
	
	\subsection{Simulation Setup}
	We consider a 3D ISAC scenario as shown in Fig.~\ref{coordinate}. The transmitted data symbols are modulated by quadrature phase shift keying (QPSK) for the fixed site scenario, and mapping through the SCMA codebook \cite{SCMA_codebook} for the multi-UE scenario. Without loss of generality, the setup given by Table \ref{table} is used.
	
	\begin{table}[h!]
		\renewcommand\arraystretch{1.3}
		\begin{center}
			\caption{Parameters used in simulations}
			\vspace{3mm}
			\label{tab1}
			\begin{tabular}{ | m{4cm} <{\centering}| m{3.8cm}<{\centering} |}
				\hline 
				\label{table}
				\textbf{Parameters} &\textbf{Values}  \\  \hline
				Location coordinate of the BS & $\mathbf{q}_B=[0,0,0]$ \\ \hline
				Location coordinates of two RISs & $\mathbf{q}_{\text{R}1}=[-30,28,21]$, $\mathbf{q}_{\text{R}2}=[-20,30,20]$ \\ \hline
				Number of UEs & $K=6$ \\ \hline
				Location coordinate ranges of $K$ UEs & $\mathbf{p}_{\text{U}k,\text{x}} \in [10,30], \  \mathbf{p}_{\text{U}k,\text{y}} \in [5,25], \ \mathbf{p}_{\text{U}k,\text{z}} \in [0,20]$  \\ \hline
				Location coordinate of the fixed site &   $\mathbf{p}_{\text{U}0}=[10,20,5]$  \\ \hline
				Number of BS's antennas &   $M=16$   \\ \hline
				Number of RIS's element &   $N_1=N_2=12\times12$    \\ \hline
				Number of paths between the BS and RISs  &   $L=3$    \\ \hline
				Number of subcarriers & $N_\text{s}=4$ \\ \hline
				Number of transmitted symbols & $T_1=T_2=256$ \\ \hline
				Power factor of data in superimposed symbols & $\xi_0=\xi_k=0.5, \ \forall k$ \\ \hline
				Values set in the proposed algorithms & $\varrho=1$, $\delta_{\omega}=0.1$, $\delta_{\chi} = \delta_{\psi} = \delta_{\eta} = 10^{-3}$, $\text{G}_1=\text{G}_2=\text{G}_3=8$, $\text{G}_4=\text{G}_5=16$ \\ \hline
			\end{tabular}
		\end{center}
	\end{table}
	
	\subsection{Performance Evaluation in the Fixed Site Scenario}
	In this part, we focus on the channel angle sensing of BS-RIS links and communication performance for the fixed site scenario. To evaluate the sensing performance, we define NMSE$(\mathbf{H}_r)$ and NMSE$(\boldsymbol{\phi})$ as the normalized mean square error (NMSE) in channel estimation and scatter angle estimation, respectively. Suppose that $\hat{\mathbf{H}}_{\text{r},i}$ and $\hat{\boldsymbol{\phi}_i}$ are the estimates of true value $\mathbf{H}_{\text{r},i}$ and $\boldsymbol{\phi}_i$, respectively, where $\boldsymbol{\phi}_i = [u_{\text{\text{R2B}},il}^{\text{A}}, u_{\text{\text{R2B}},il}^{\text{D}}, v_{\text{\text{R2B}},il}^{\text{D}}]_{l=1}^L$, $\text{NMSE}(\mathbf{H}_r) = \frac{\sum_{i=1}^2\|\mathbf{H}_{\text{r},i}-\hat{\mathbf{H}}_{\text{r},i}\|^2}{\sum_{i=1}^2\|\mathbf{H}_{\text{r},i}\|^2}$ and $\text{NMSE}(\boldsymbol{\phi}) = \frac{\sum_{i=1}^2\|\boldsymbol{\phi}_{i}-\hat{\boldsymbol{\phi}}_{i}\|^2}{\sum_{i=1}^2\|\boldsymbol{\phi}_i\|^2}$. To evaluate the communication performance, we calculate the effective SINR at the $t$-th time slot considering the estimation error. Suppose that $\hat{\mathbf{h}}_{\text{casc},t}$ is the estimate of cascaded channel at the $t$-th time slot, i.e., $\hat{\mathbf{h}}_{\text{casc},t} = \sum_{i=1}^2 \mathbf{H}_{\text{r},i} \text{diag}(\boldsymbol{\theta}_{i,t}) \mathbf{\hat{h}}_{\text{b},i0}$. The effective SINR for $t$-th time slot is given by $\text{SINR}_{0,\text{eff}}(t) = \frac{\xi_0  \|\mathbf{\hat{h}}_{\text{casc},t}\|^2}{(1-\xi_0)\|(\mathbf{{h}}_{\text{casc},t}-\mathbf{\hat{h}}_{\text{casc},t})\|^2+N_0}$.
	
	\subsubsection{ISAC performance comparison with existing algorithms}
	\begin{figure*}
		\begin{center}
			\subfigure[BER performance comparison against the SNR.]{
				\label{benchmark1_BER}
				\includegraphics[width=3.4in]{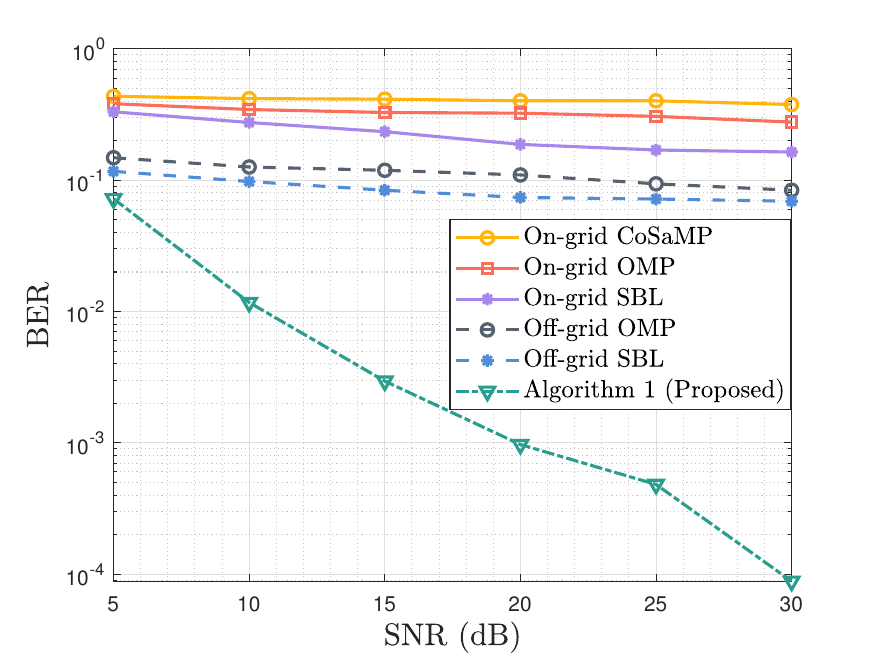}
			}
			\subfigure[NMSE of channel estimation comparison against SNR.]{
				\label{benchmark1_NMSE}
				\includegraphics[width=3.2in]{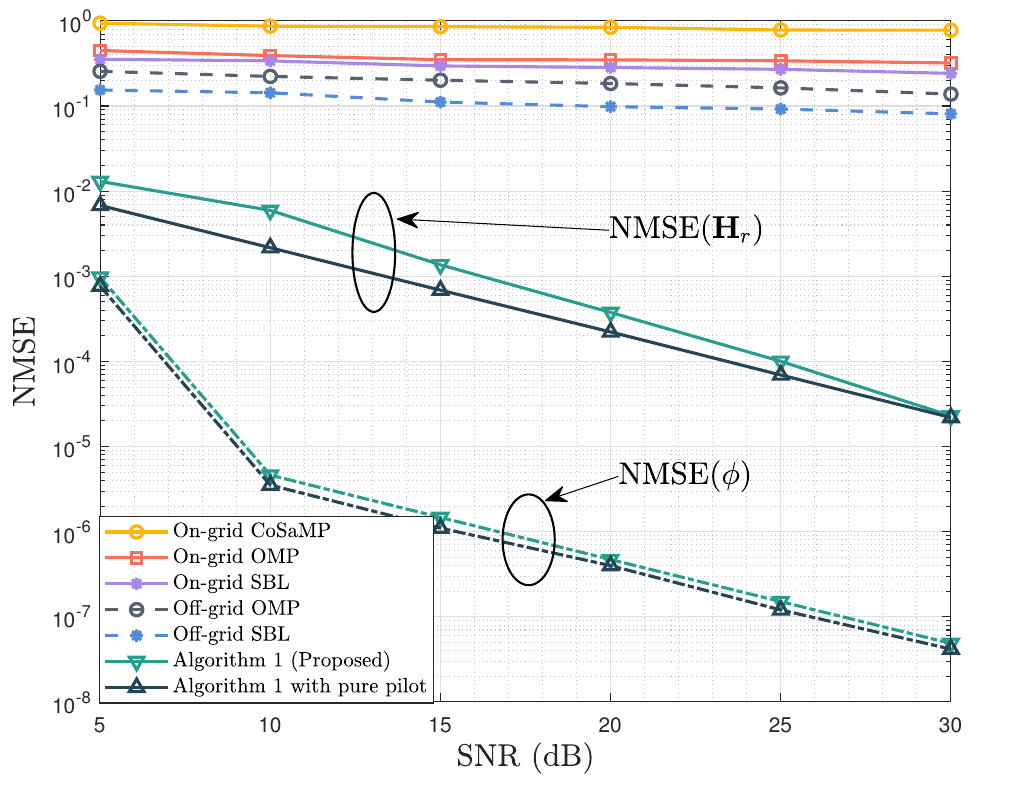}
			}
			\caption{ISAC performance comparison with existing algorithms.}
			\label{benchmark1}
		\end{center}
	\end{figure*}
	Fig.~\ref{benchmark1} shows the superior performance of our proposed \emph{Algorithm 1} on BER, NMSE$(\mathbf{H}_r)$ and NMSE$(\boldsymbol{\phi})$ versus different SNR regimes. For comparison, we use three on-grid based methods and two off-grid based methods, all of which update $\boldsymbol{\omega}$ without iterative procedures. We first compare the three on-grid based methods with \emph{off-grid SBL} and \emph{off-grid OMP} in Fig.~\ref{benchmark1_BER}, which shows that the severe model mismatch by on-grid methods heavily degrades ISAC performance. Furthermore, it can be observed that even using off-grid based methods, i.e., \emph{off-grid SBL} and \emph{off-grid OMP}, provides a small improvement in terms of NMSE and BER since the symbols used for sensing contain unknown data. Specifically, the BER of these unknown data is nearly $10^{-1}$, which in turn, reduces sensing capabilities. This highlights the benefits of updating data symbols via an iterative manner as in \emph{Algorithm 1}, especially in the high SNR range. In addition, the performance of the \emph{Algorithm 1 with pure pilot} is included as a lower bound for the proposed superimposed symbol scheme in Fig.~\ref{benchmark1_NMSE}. The NMSE of \emph{Algorithm 1} is very close to this lower bound, indicating that the benefits of the proposed algorithm are comparable to removing most of the interference between sensing pilots and data, especially at high SNR.
	
	\subsubsection{SE and NMSE performance by various transmission schemes}
	\begin{figure*}
		\begin{center}
			\subfigure[SE performance comparison against SNR.]{
				\label{stage1_SE}
				\includegraphics[width=3.3in]{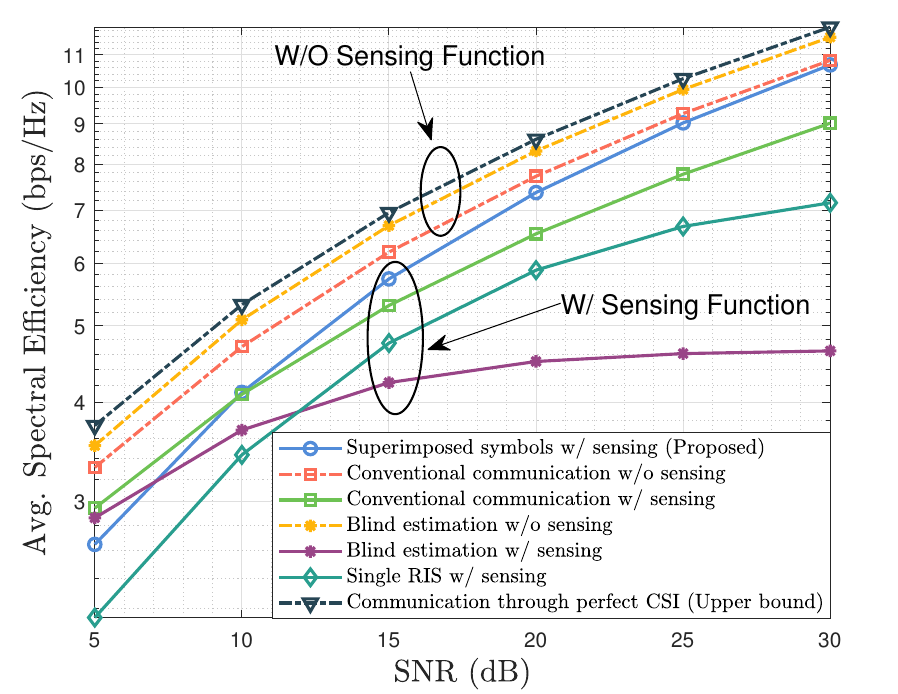}
			}
			\subfigure[NMSE performance comparison against SNR.]{
				\label{stage1_SE_NMSE}
				\includegraphics[width=3.4in]{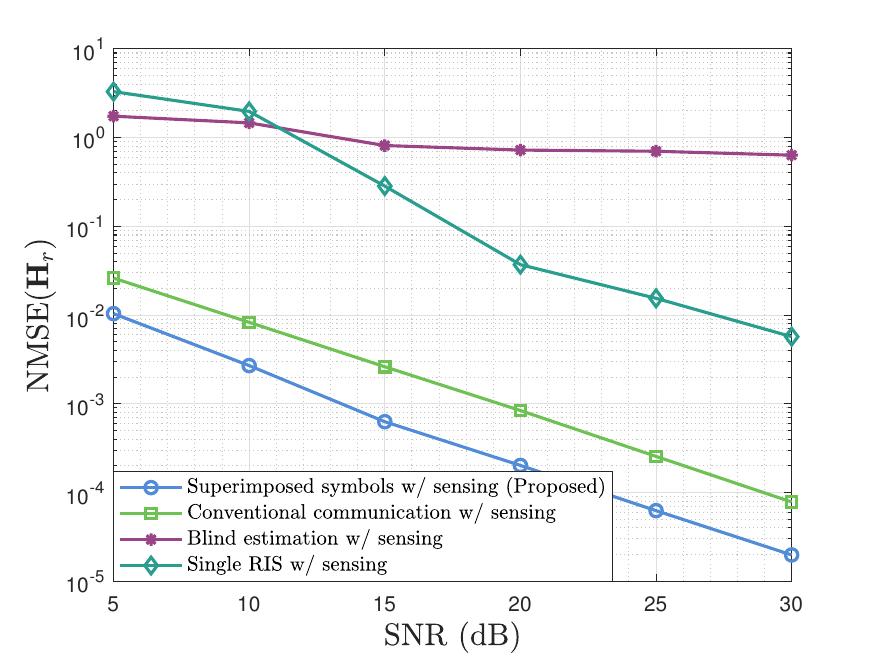}
			}
			\caption{ISAC performance comparison with various transmission schemes.}
			\label{SE}
		\end{center}
	\end{figure*}
	Results in Fig.~\ref{stage1_SE} underline the valuable role of the proposed superimposed symbol scheme in compensating for the SE degradation caused by having sensing capabilities, where we consider two cases, i.e., with and without sensing functions denoted by \emph{``W/ Sensing Function''} and \emph{``W/O Sensing Function''}, respectively. It can be observed that the addition of a sensing function lowers the SE, as more sensing pilots consume spectrum resources. Surprisingly, the proposed superimposed symbol scheme provides comparable SE to that of conventional communication without sensing function in the high or moderate SNR regime, achieving $96\%$ SE at $20$ dB. This is because our proposed superimposed symbol scheme can obtain accurate channel angle information which can nearly eliminate the mutual interference between pilots and data. Moreover, we also evaluate the typical blind estimation with very few pilots, which is shown as the solution closest to the upper bound through perfect CSI. However, once the blind estimation method is implemented to have the sensing function, SE will be severely degraded due to the demand for extensive pilots to eliminate the quadrant ambiguity in the channel angle extraction procedure. In addition, compared with the double-RIS setup, the single-RIS scenario still experiences a decrease in SE, even though it does not need to deal with the interference between double RISs. This is because the other RIS is turned OFF, resulting in a decrease in channel gain.
	
	Fig.~\ref{stage1_SE_NMSE} illustrates the NMSE performance among the abovementioned methods with the sensing function. The numerical results show again that the proposed superimposed symbol scheme can significantly outperform other benchmarks. This trend is almost in agreement with the SE performance. The reason for this behavior is that the method with high SE can yield more data symbols, which in turn improves the channel sensing performance.

	\subsubsection{Impact of the transmitted symbols length $T_1$}
	\begin{figure*}
		\begin{center}
			\subfigure[NMSE of channel estimation comparison against $T_1$.]{
				\label{NMSE_T1}
				\includegraphics[width=3.8in,height=2in]{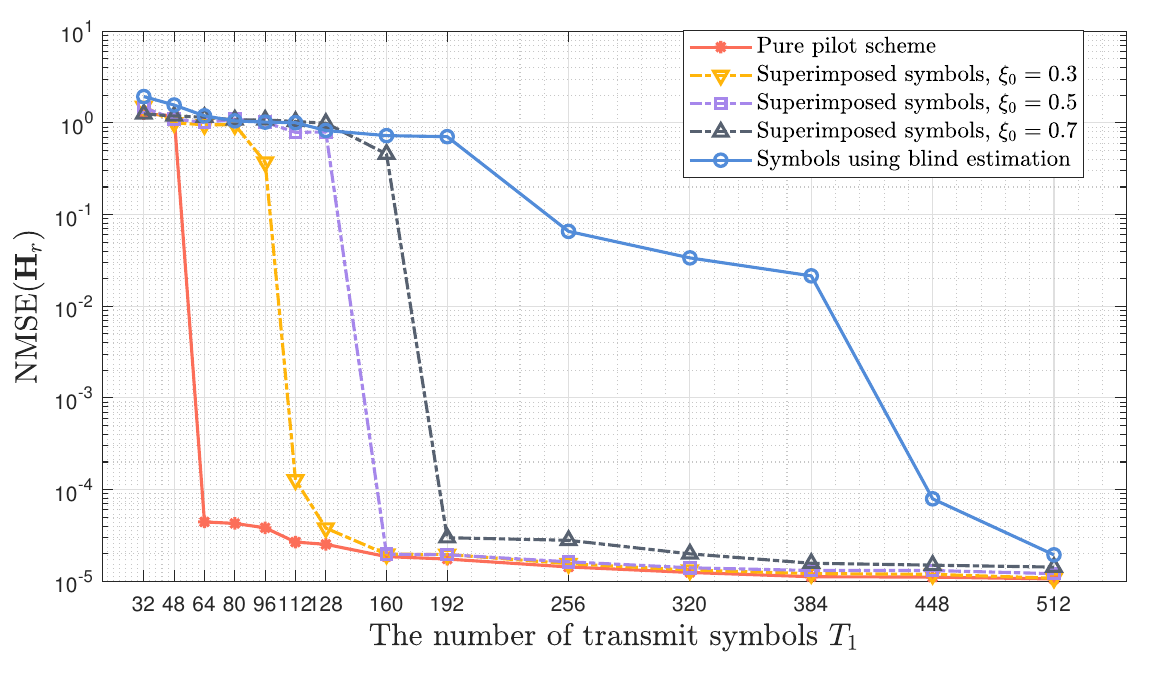}
			}
			\subfigure[Effective throughput amount comparison against $T_1$.]{
				\label{eff_T1}
				\includegraphics[width=2.8in]{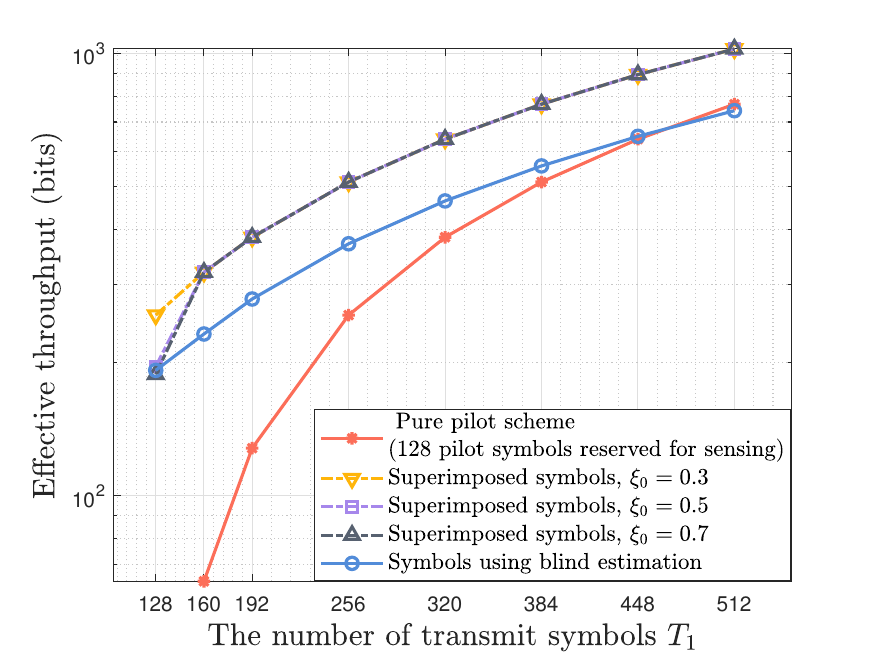}
			}
			\caption{Impact of the length of the transmitted symbol on the ISAC performance.}
			\label{T1}
		\end{center}
	\end{figure*}
	Fig.~\ref{NMSE_T1} exhibits how many symbols are needed to achieve satisfactory sensing performance at SNR=$30$ dB. It shows that the NMSE is a decreasing function of $T_1$ for all methods and eventually reaches stable accuracy. Regarding the cases of superimposed symbol schemes, abrupt transitions around $T_1$ = $112$, $160$, and $192$ can be observed in the fast fading regime, where $T_1$ threshold values increase with $\xi_0$. This can be explained by the fact that more symbols are needed to reduce NMSE while more power is allocated to data.    
	
	Fig.~\ref{eff_T1} illustrates the performance gain in effective throughput by our proposed superimposed symbol scheme. It shows that the superiority of the superimposed symbol scheme covers the entire $T_1$ range. This effect is explained by the fact that sensing pilots are inserted into each transmission block along with data symbols, resulting in little dedicated consumption of time-frequency resources for communications. Particularly, the effective throughput by the superimposed symbol scheme is $200\%$ larger than that of the pilot-based sensing scheme and $133\%$ larger than the blind-based sensing scheme at $T_1 = 256$.

	\subsection{ISAC Performance Evaluation for the Multiple UEs}
	In this part, we provide numerical results to verify the effectiveness of the proposed algorithm for multi-UE simultaneous communication and localization. To show the localization performance, we first use the localization error to evaluate the average localization accuracy, defined as $\frac{1}{K} \sum_{k=1}^K \|\hat{\mathbf{p}}_{\text{U},k} - \mathbf{p}_{\text{U},k}\|$, where $\hat{\mathbf{p}}_{\text{U},k}$ is the localization estimate for the $k$-th UE. To show the communication performance, we derive the effective SINR for the $k$-th UE at the $t$-th time slot considering the channel estimation error as
	$\text{SINR}_{k,\text{eff}}(t) = \\	
	\frac{\sigma_{k,d}^2 \|\mathbf{\hat{h}}_{\text{cast},k,t}\|^2}{ \|(\mathbf{{h}}_{\text{cast},k,t}-\mathbf{\hat{h}}_{\text{cast},k,t}) x_{k,t}\|^2+ \sum\limits_{q \ne k}^K \sigma_{q,d}^2 \| \mathbf{\hat{h}}_{\text{cast},k,t}\|^2+\|(\mathbf{{h}}_{\text{cast},q,t}-\mathbf{\hat{h}}_{\text{cast},q,t}) x_{q,t}\|^2}$, where $\mathbf{\hat{h}}_{\text{cast},k,t}$ is the estimate of the cascaded channel for the $k$-th UE and $t$-th time slot, where $\mathbf{\hat{h}}_{\text{cast},k,t} =  \sum_{i=1}^2 \mathbf{H}_{\text{r},i} \text{diag}(\boldsymbol{\theta}_{i,t}) \mathbf{\hat{h}}_{\text{b},ik}$ and $\mathbf{\hat{h}}_{\text{b},ik}$ is the estimate of $\mathbf{{h}}_{\text{b},ik}$.
	
	\subsubsection{Localization and SE performance with different multi-access schemes}
	\begin{figure*}
		\begin{center}
			\subfigure[Localization performance comparison against $E_b/N_0$.]{
				\label{SCMA}
				\includegraphics[width=3.4in]{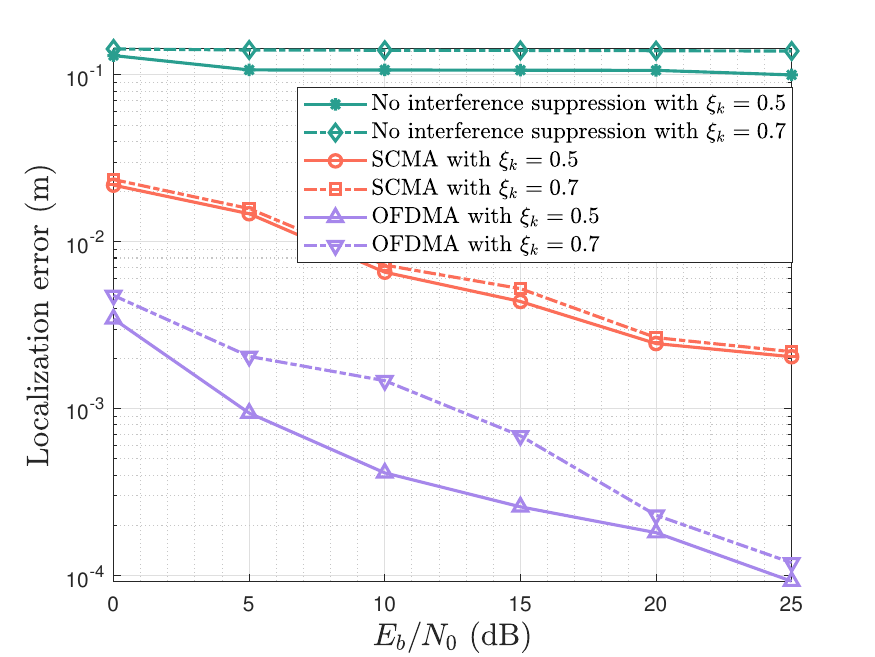}
			}
			\subfigure[SE performance comparison against $E_b/N_0$.]{
				\label{SCMA_SE}
				\includegraphics[width=3.4in]{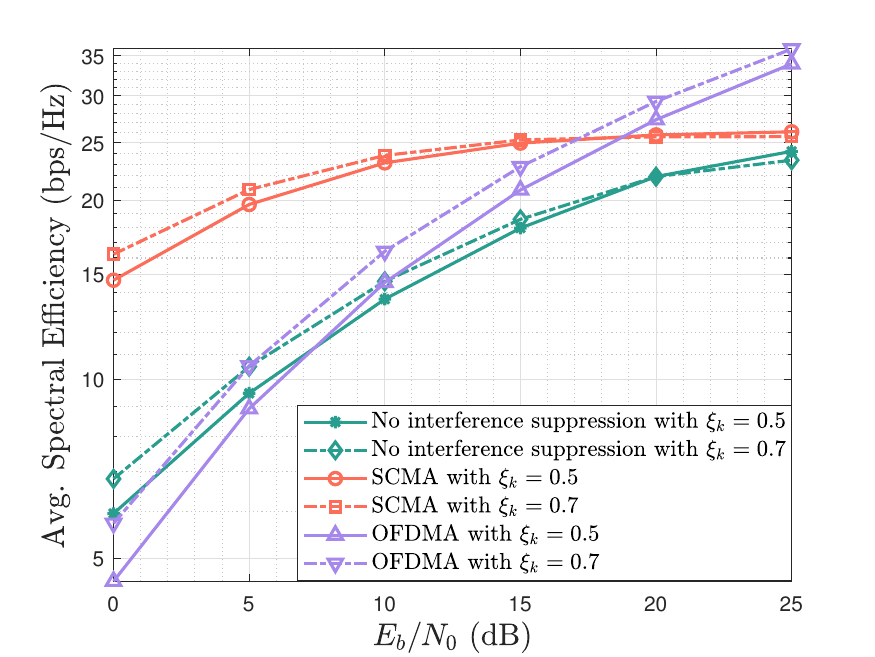}
			}
			\caption{ISAC performance comparison with benchmarks.}
			\label{SCMA1}
		\end{center}
	\end{figure*}
	
	Fig.~\ref{SCMA1} illustrates the impact of using different multi-access schemes on the ISAC performance. We adopt the orthogonal frequency division multiple access (OFDMA) scheme with $K=6$ and $N_\text{s}=6$ and no interference suppression case with $K=6$ and $N_\text{s}=3$ for comparison. As shown in Fig.~\ref{SCMA}, OFDMA modulation has the optimal localization performance, however, frequency resources are highly demanded, especially when there is a large number of UEs. As a preferred solution considering both resources and ISAC performance, the proposed SCMA scheme can achieve a localization accuracy of $10^{-2}$ m at 10 dB and improves significantly with increasing transmit power. However, the localization accuracy can only reach $0.1$ m when there is no proper inter-UE interference suppression strategy.
	
	In Fig.~\ref{SCMA_SE}, we can see that the SCMA scheme achieves the best SE performance at moderate SNR regimes since it requires fewer resources to achieve reliable communications. Even though the OFDMA scheme has the best SE performance at high SNR, the proposed SCMA scheme is suitable for serving more UEs when resources are limited. Moreover, it can be observed that $\xi_k$ can also significantly affect the ISAC performance. Therefore, selecting the optimal $\xi_k$ is an important issue, which can be solved through numerical results and omitted in this paper for brevity.

	\subsubsection{Impact of the transmitted symbols length $T_2$}
\begin{figure}[ht]
		\centering
		\includegraphics[scale=0.6]{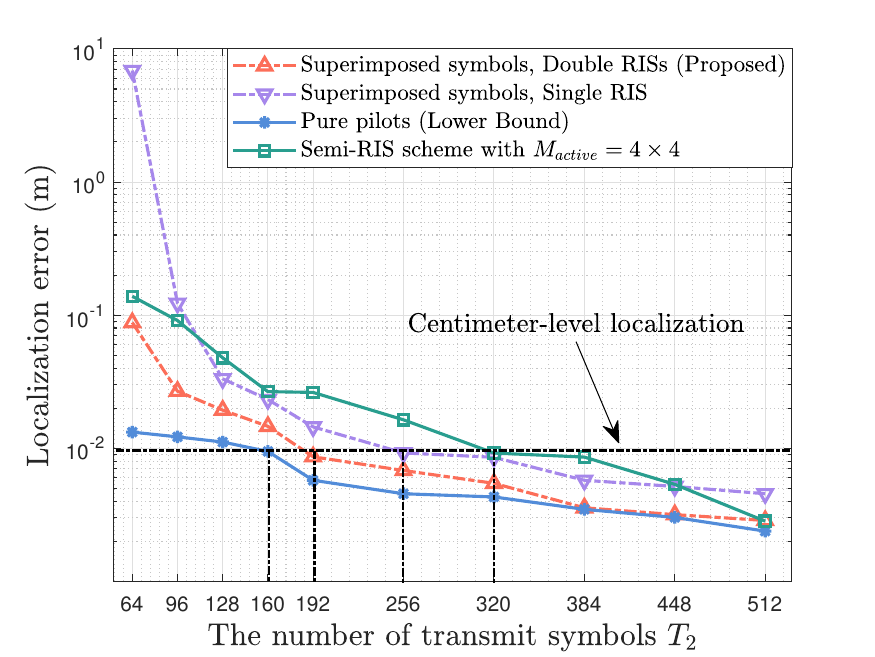}
		\caption{Localization performance comparison against $T_2$.}
		\label{loc_T2}
	\end{figure}

	Fig.~\ref{loc_T2} exhibits the localization performance of different methods against various $T_2$ at $15$ dB. The semi-RIS scheme \cite{hu_tsp} with $M_{\text{active}}=M$ is adopted as a benchmark under fair comparisons. It can be observed that all these schemes can achieve centimeter-level localization, but the demands of symbol lengths are varying. More specifically, the pure pilot scheme requires $160$ symbols, while the proposed superimposed symbol scheme only needs $192$ symbols, which is lower than other schemes. This indicates that the proposed scheme requires fewer symbol resources to realize communication capabilities while achieving high-accuracy localization as well.
	
	\section{CONCLUSIONS}\label{conclusion}
	In this paper, we proposed a superimposed symbol scheme for double-RIS aided ISAC mmWave systems to achieve simultaneous multi-UE communication and localization. Specifically, the structure-aware SBL method was designed to enable simultaneous communication and channel angle sensing. Considering the low-latency requirements in ISAC applications, a low-complexity algorithm for multi-UE ISAC scenarios was further developed, where the UAMP-SBL method was utilized to obtain initial channel angle estimates which were then refined iteratively by reduced-dimension matrix calculations. Besides, the SCMA technique is incorporated into this iterative framework for accurate data detection and localization. Simulation results have demonstrated that our proposed superimposed symbol-based scheme empowered by the developed algorithm can provide centimeter-level localization accuracy with $96\%$ SE of conventional communications without sensing function. This work has verified the effectiveness and immense potential of using the superimposed symbol scheme in ISAC systems in the Bayesian learning framework. Furthermore, investigating the optimization of the phase design and placement of RISs in double-RIS aided ISAC systems presents another direction for future exploration. These extensions hold the potential to enhance the overall ISAC performance in diverse and expansive scenarios.

	\appendices
	\section{Derivation of $\beta^{(j-1)}$ and $\boldsymbol{\nu}^{(j-1)}$}\label{beta_nu}
	The maximization step of the first term related to $\beta$ can be expressed as 
	\begin{equation}\label{beta}
		\begin{aligned}
			&\beta^{(j)} = \arg \max_{\beta} MT_1 \ln(\beta) -\beta \big(\|\mathbf{y}_0 - \mathbf{Z}_{\text{\text{R2B}}}(\mathbf{X}_0^{(j)}, \boldsymbol{\nu}^{(j-1)}) \\
			&\boldsymbol{\mu}_{\boldsymbol{\omega}}^{(j-1)}\|^2 \!\! +\!\text{Tr}( \mathbf{Z}_{\text{\text{R2B}}}(\mathbf{X}_0^{(j)}\!, \boldsymbol{\nu}^{(j-1)}) \boldsymbol{\Sigma}_{\boldsymbol{\omega}}^{(j-1)} \mathbf{Z}_{\text{\text{R2B}}}^H(\mathbf{X}_0^{(j)}\!, \boldsymbol{\nu}^{(j-1)})  ) \big),
		\end{aligned}
	\end{equation}
	which is a concave function w.r.t $\beta$, setting its derivative to zero gives the unique optimal solution as in (\ref{beta^j}). Then, the objective function in terms of $\boldsymbol{\nu}_2$ is reduced to
	\begin{equation}\label{nu}
		\begin{aligned}
			\boldsymbol{\nu}_2^{(j)} 
			& = \arg \max_{\boldsymbol{\nu}_2}  \mathbb{E}_{\boldsymbol{\omega} | \mathbf{y}_0 ; \boldsymbol{\chi}^{(j-1)}} - \beta^{(j)} \bigg(\|\mathbf{y}_0 - \mathbf{Z}_{\text{\text{R2B}}}(\mathbf{X}_0^{(j)}, \boldsymbol{\nu}_2) \\
			&\boldsymbol{\mu}_{\boldsymbol{\omega}}^{(j-1)}\|^2 +\text{Tr}\left( \mathbf{Z}_{\text{\text{R2B}}}(\mathbf{X}_0^{(j)}, \boldsymbol{\nu}_2) \boldsymbol{\Sigma}_{\boldsymbol{\omega}}^{(j-1)} \mathbf{Z}_{\text{\text{R2B}}}^H(\mathbf{X}_0^{(j)}, \boldsymbol{\nu}_2)  \right) \bigg)
		\end{aligned}
	\end{equation}
	
	However, it is not easy to obtain a closed-form solution to (\ref{nu}) since the problem is highly non-linear in $\boldsymbol{\nu}$. Therefore, we search for the updated $\boldsymbol{\nu}_2^{(j)}$ via sub-gradient method \cite{stepsize} as
	\begin{equation}
		\boldsymbol{\nu}_2^{{(j)}} = \boldsymbol{\nu}_2^{(j-1)} + \epsilon_1^{(j)}  \mathcal{F}_1(\boldsymbol{\nu}_2^{(j-1)})
	\end{equation}
	where $\epsilon_1^{(j)}$ is the stepsize selected for the $j$-th iteration and $\mathcal{F}_1(\boldsymbol{\nu}_2^{(j-1)})$ is the derivative of (\ref{nu}). The derivative $\mathcal{F}_1(\boldsymbol{\nu}_2^{(j-1)})$ can be calculated in a closed-form and omitted here. Moreover, we can resort to backtracking line search \cite{line_search} to choose the appropriate stepsize, which ensures that the objective value can be strictly decreased before reaching the stationary point. However, the complexity of each update stepsize needs to be reduced \cite{off_liu}. Then, we alternatively use a certain stepsize with learning rate decay to achieve arbitrarily small precision stepsizes after sufficient iterations\cite{line_search}, expressed as
	\begin{equation}\label{epsilon1}
		\epsilon_1^{(j)} = \frac{\nu_{\max}- \nu_{\min}}{c_1 (1+d_1 \cdot j)}
	\end{equation}  
	where $\nu_{\max}$ and $\nu_{\min}$ are the maximum and minimum values of the selected girds, respectively. $c_1$ and $d_1$ are learning rate and decay coefficients, which are required to be carefully selected to achieve the optimal performance. Furthermore, the proof of convergence of the sub-gradient descent method can refer to the work \cite{boyd}.

	\section{The Derivation of \emph{Lemma 1}}\label{X_d,0}
	The first term of (\ref{li}) can be transformed as $\mathbb{E}_{\mathbf{H}_{\text{eff}}| \mathbf{Y}_0; \boldsymbol{\chi}^{(j-1)}} \left\{ \ln \left[ p(\mathbf{Y}_0| \mathbf{H}_{\text{eff}}; \mathbf{X}_{\text{d},0}, \boldsymbol{\nu}^{(j-1)}, \beta^{(j-1)}) \right] \right\}$ to investigate the linear relationships, where $\mathbf{H}_{\text{eff}} = \sum_{i=1}^2  \sqrt{\frac{M N_i}{L_i}} \mathbf{A}_{\text{\text{R2B}},i} \text{vec}^{-1}(\boldsymbol{\omega}_i) \mathbf{B}_{\text{\text{R2B}},i}^H \text{diag}(\mathbf{h}_{i,0}) \boldsymbol{\Theta}_i$. Then the maximization step of (\ref{li}) in terms of $\mathbf{X}_{\text{d},0}$ can be expressed as
	\begin{equation}\label{X}
		\begin{aligned}
			&\arg \max\limits_{\mathbf{X}_{\text{d},0}} \mathbb{E}_{\mathbf{H}_{\text{eff}}| \mathbf{Y}_0; \boldsymbol{\chi}^{(j-1)}} \left\{ \ln \left[ p(\mathbf{Y}_0| \mathbf{H}_{\text{eff}}; \mathbf{X}_{\text{d},0}, \boldsymbol{\nu}^{(j-1)}, \beta^{j-1}) \right] \right\} \\
			& = \arg \min\limits_{\mathbf{X}_{\text{d},0}} \text{Tr} \bigg\{  -2\sqrt{\xi_0}\text{Re}[ \mathbf{Y}_0^H \mathbb{E}\{\mathbf{H}_{\text{eff}}\} 
			\mathbf{X}_{\text{d},0} ] +\xi_0\mathbf{X}_{\text{d},0}^H\cdot \\
			& \mathbb{E}\{\mathbf{H}_{\text{eff}}^H \mathbf{H}_{\text{eff}}\} \mathbf{X}_{\text{d},0} + 2 \sqrt{\xi_0 (1-\xi_0)}\text{Re}\big[ \mathbf{X}_{\text{d},0}^H \mathbb{E}\{\mathbf{H}_{\text{eff}}^H \mathbf{H}_{\text{eff}}\} \mathbf{X}_{\text{p},0}\big] \bigg\}.
		\end{aligned}
	\end{equation}
	
	Through the affine transformation, $p(\mathbf{H}_{\text{eff}} | \mathbf{Y}_0; \boldsymbol{\chi}^{(j-1)})$ also obeys the complex Gaussian distribution $\mathcal{CN}(\boldsymbol{\mu}_{H\text{eff}}^{(j-1)},\boldsymbol{\Sigma}_{H\text{eff}}^{(j-1)})$. As such, we have $\mathbb{E}\{\mathbf{H}_{\text{eff}}\} =  \boldsymbol{\mu}_{H\text{eff}}^{(j-1)}$ and $\mathbb{E}\{\mathbf{H}_{\text{eff}}^H \mathbf{H}_{\text{eff}}\} = \left(\boldsymbol{\mu}_{H\text{eff}}^{(j-1)}\right)^H \boldsymbol{\mu}_{H\text{eff}}^{(j-1)} + \boldsymbol{\Sigma}_{H\text{eff}}^{(j-1)}$, where
	\begin{equation}\label{mu_H}
		\begin{aligned}
			\boldsymbol{\mu}_{H\text{eff}}^{(j-1)} =\sum\limits_{i=1}^2& \sqrt{\frac{M N_i}{L_i}} \mathbf{A}_{\text{\text{R2B}},i} \left[\text{vec}^{-1}\left(\boldsymbol{\mu}_{\boldsymbol{\omega}_i}^{(j-1)} \right)  \right] \mathbf{B}_{\text{\text{R2B}},i}^H \\
			& \cdot \text{diag}(\mathbf{h}_{\text{b},i0}) \boldsymbol{\Theta}_i
		\end{aligned}
	\end{equation}
	
	Denoting $\mathbf{h}_\text{eff} = \text{vec}(\mathbf{{H}}_\text{eff})$ with the associated covariance matrix $\boldsymbol{\Sigma}_{h\text{eff}}^{(j-1)}$, we have $\boldsymbol{\Sigma}_{h\text{eff}}^{(j-1)} = \boldsymbol{\Phi} \boldsymbol{\Sigma}_{\boldsymbol{\omega}}^{(j-1)} \boldsymbol{\Phi}^H$, where $\boldsymbol{\Phi} = [
	\sqrt{\frac{M N_1}{L_1}} \left( \mathbf{B}_{\text{\text{R2B}},1}^H \text{diag}(\mathbf{h}_{\text{b},10}) \boldsymbol{\Theta}_1  \right)^T \otimes \mathbf{A}_{\text{\text{R2B}},1} , \ \sqrt{\frac{M N_2}{L_2}} \\ \left( \mathbf{B}_{\text{\text{R2B}},2}^H \text{diag}(\mathbf{h}_{\text{b},20}) \boldsymbol{\Theta}_2  \right)^T \otimes \mathbf{A}_{\text{\text{R2B}},i},
	]$. Then, we have
	\begin{equation}\label{sigma_H}
		\begin{aligned}
			&\boldsymbol{\Sigma}_{H\text{eff}}^{(j-1)}(t_i,t_j) =\\
			& \ \text{Tr} \left\{\boldsymbol{\Sigma}_{h\text{eff}}^{(j-1)}\left((t_i-1)M+1:t_iM, (t_j-1)M+1:t_jM \right)  \right\} ,
		\end{aligned}
	\end{equation}
	which obeys the MATLAB writing format and $1\leq t_i,t_j \leq T_1$. Then, setting the first-order derivative of (\ref{X}) to zero yields the result in Eq.~(\ref{xd}).
	
	\section{The Derivation of $\mathbf{\hat{h}}_{\text{b},ik}^{(0)}$}\label{h^0}
	Considering the vectorization $\mathbf{y}_{\text{U},n_\text{s}} = \text{vec}(\mathbf{Y}_{\text{U},n_\text{s}})$ in Eq.~(\ref{YU}), and employing the property $\text{vec}(\mathbf{A} \text{diag}(\mathbf{b}) \mathbf{C}) = (\mathbf{C}^T \odot \mathbf{A}) \mathbf{b}$, where $\odot$ represents the Khatri-Rao product, we have
	\begin{equation}
		\mathbf{y}_{\text{U},n_\text{s}} = \sum\limits_{k=1}^K \sum\limits_{i=1}^2 \left( (\boldsymbol{\Theta}_i \mathbf{X}_{k,n_\text{s}} )^T \odot \mathbf{H}_{\text{r},i} \right) \mathbf{h}_{\text{b},ik} + \mathbf{n}_{\text{U},n_\text{s}},
	\end{equation}
	
	By introducing the mmWave sparse model of $\mathbf{h}_{\text{b},ik}$ in Eq.~(\ref{hb_sparse}), we have
	\begin{equation}\label{yU_a}
		\mathbf{y}_{\text{U},n_\text{s}} = \mathbf{V}(\mathbf{X}_{n_\text{s}}) \mathbf{B}_{\text{U}2}(\mathbf{w}_{\text{\text{U2R}},i}^{\text{A}}, \mathbf{g}_{\text{\text{U2R}},i}^{\text{A}}) \boldsymbol{\psi}+ \mathbf{n}_{\text{U},n_\text{s}},
	\end{equation}
	where $\mathbf{V}(\mathbf{X}_{n_\text{s}}) = \big[\mathbf{V}_1(\mathbf{X}_{1,n_\text{s}}),...,\mathbf{V}_K(\mathbf{X}_{K,n_\text{s}})  \big]\in  \mathbb{C}^{MT_2\times 2KN}$, $\mathbf{V}_k(\mathbf{X}_{k,n_\text{s}}) = [\sqrt{N_1} \left( (\boldsymbol{\Theta}_1 \mathbf{X}_{k,n_\text{s}} )^T \odot \mathbf{H}_{\text{r},1} \right) \\, \sqrt{N_2} ( (\boldsymbol{\Theta}_2 \mathbf{X}_{k,n_\text{s}} )^T \odot \mathbf{H}_{\text{r},2} ) ]$, $\mathbf{B}_{\text{U}2}(\mathbf{w}_{\text{\text{U2R}},i}^{\text{A}}, \mathbf{g}_{\text{\text{U2R}},i}^{\text{A}}) = \text{blkdiag}\big{\{}  \mathbf{B}_{\text{U}}(\mathbf{w}_{\text{\text{U2R}},i}^{\text{A}}, \mathbf{g}_{\text{\text{U2R}},i}^{\text{A}}),..., \mathbf{B}_{\text{U}}(\mathbf{w}_{\text{\text{U2R}},i}^{\text{A}}, \mathbf{g}_{\text{\text{U2R}},i}^{\text{A}})\big{\}} \\ \in \mathbb{C}^{2KN \times 2K\text{G}_4\text{G}_5}$, $\mathbf{B}_{\text{U}}(\mathbf{w}_{\text{\text{U2R}},i}^{\text{A}}, \mathbf{g}_{\text{\text{U2R}},i}^{\text{A}}) = \text{blkdiag}\big{\{}  \mathbf{B}_{\text{\text{U2R}},1}(\mathbf{w}_{\text{\text{U2R}},1}^{\text{A}}, \mathbf{g}_{\text{\text{U2R}},1}^{\text{A}}), \ \mathbf{B}_{\text{\text{U2R}},2}(\mathbf{w}_{\text{\text{U2R}},2}^{\text{A}}, \mathbf{g}_{\text{\text{U2R}},2}^{\text{A}})\big{\}} $, blkdiag$(\cdot)$ is the block diagonalization operation, $\boldsymbol{\psi} = [\boldsymbol{\psi}_1^T,...,\boldsymbol{\psi}_K^T]^T$ and $\boldsymbol{\psi}_k = [\boldsymbol{\psi}_{1,k}^T, \ \boldsymbol{\psi}_{2,k}^T]^T$.
	
	Then, we collect (\ref{yU_a}) in $N_\text{s}$ frequency bands to form
	\begin{equation}\label{yU2_a}
		\mathbf{y}_U = \mathbf{V}_2(\mathbf{X}) \mathbf{B}_{\text{U}2}(\mathbf{w}_{\text{\text{U2R}},i}^{\text{A}}, \mathbf{g}_{\text{\text{U2R}},i}^{\text{A}}) \boldsymbol{\psi}+ \mathbf{n}_{\text{U}},
	\end{equation}
	where $\mathbf{y}_{\text{U}} = [\mathbf{y}_{\text{U},1}^T,...,\mathbf{y}_{\text{U},N_\text{s}}^T]^T \in \mathbb{C}^{MT_2N_\text{s} \times 1}$, $\mathbf{V}_2(\mathbf{X}) = [\mathbf{V}^T(\mathbf{X}_1),...,\mathbf{V}^T(\mathbf{X}_{N_\text{s}})]^T\in \mathbb{C}^{MT_2N_\text{s} \times 2KN}$ and $\mathbf{n}_U = [\mathbf{n}_{\text{U},1}^T,...,\mathbf{n}_{\text{U},N_\text{s}}^T]^T$.   
	
	By converting the received signal into the form as Eq.~\eqref{yU2_a} and substituting $\sqrt{1-\xi} \mathbf{X}_p$ for $\mathbf{X}$, we can use LSE method to directly obtain $\boldsymbol{\psi}$ as
	\begin{equation}
		\boldsymbol{\psi}^{(0)} = \left[ \mathbf{V}_2(\mathbf{X}) \mathbf{B}_{\text{U}2}(\mathbf{w}_{\text{\text{U2R}},i}^{\text{A}}, \mathbf{g}_{\text{\text{U2R}},i}^{\text{A}}) \right]^{-1} \mathbf{y}_U
	\end{equation}
	
	In this way, the intial estimation of $\mathbf{\hat{h}}_{\text{b},ik}^{(0)}$ can be expressed as
	\begin{equation}
		\mathbf{\hat{h}}_{\text{b},ik}^{(0)} = \sqrt{N_i} \mathbf{B}_{\text{\text{U2R}},i}(\mathbf{w}_{\text{\text{U2R}},i}^{\text{A}}, \mathbf{g}_{\text{\text{U2R}},i}^{\text{A}}) \boldsymbol{\psi}_{i,k}^{(0)}.
	\end{equation}

\end{document}